\documentclass[aps,pra,showpacs,amsfonts,notitlepage,amssymb,amsmath,nofootinbib,superscriptaddress,twocolumn,longbibliography]{revtex4-1}
\usepackage[utf8]{inputenc}
\usepackage[english]{babel}
\usepackage[titletoc,toc,title]{appendix}
\usepackage{amsmath}
\usepackage{amsfonts}
\usepackage{amssymb}
\usepackage[colorlinks=true,linkcolor=blue,citecolor=blue,urlcolor=blue]{hyperref}
\usepackage{graphicx}
\usepackage{enumerate}
\usepackage{csquotes}
\usepackage[caption=false]{subfig}
\usepackage{dsfont}
\usepackage{color}
\usepackage{bbold}
\usepackage{mathtools}
\usepackage{tensor}
\usepackage{gensymb}

\newcommand{\unotp}{u_0^+}

\begin{document}

\title{Vortical scattering channel in an aquatic space-time}

\author{Alessia Biondi}
\affiliation{Institut Pprime, CNRS--Université de Poitiers--ISAE-ENSMA. TSA 51124, 86073 Poitiers Cedex 9, France}
\author{Scott Robertson}
\affiliation{Institut Pprime, CNRS--Université de Poitiers--ISAE-ENSMA. TSA 51124, 86073 Poitiers Cedex 9, France}
\author{Germain Rousseaux}
\affiliation{Institut Pprime, CNRS--Université de Poitiers--ISAE-ENSMA. TSA 51124, 86073 Poitiers Cedex 9, France}

\begin{abstract}
Effective field theory descriptions of surface waves on flowing fluids have tended to assume that the flow is irrotational, but this assumption is often impractical due to boundary layer friction and flow recirculation. Here we develop an effective field theory of surface waves in an incompressible, inviscid flow that includes vorticity due to shear. Our model consists of a two-layer flow: an upper layer with no vorticity and a lower layer with constant vorticity. We consider linear, long-wavelength perturbations on top of such a flow, and find that these can be described by two coupled scalar fields admitting three elementary excitations, one more than the usual two found in irrotational flows. We compute the scattering coefficients pertaining to modes falling into an analogue black hole. Our approach provides a more realistic framework for simulating gravitational wave phenomena possibly with an internal structure mimicking quantum gravity effects in laboratory settings.
\end{abstract}

\maketitle

Surface waves on flowing fluids can be described as excitations of a scalar field propagating in an effective (2+1)-dimensional spacetime~\cite{PhysRevD.66.044019,rousseaux2013basics}.  This spacetime is described by the generalized Painlev\'{e}-Gullstrand~\cite{painleve1921mecanique} line element
\begin{equation}
    {\rm d}s^{2}  = c^{2} \left[ c^{2} \, {\rm d}t^{2} - \left( {\rm d}{\bf x}_{\parallel} - {\bf u}_{\parallel} \, {\rm d}t \right)^{2} \right] \,,
    \label{eq:PG_line_element}
\end{equation}
where ${\bf u}_{\parallel}$ is the projection of the mean flow in the horizontal plane and $c$ is the wave speed with respect to the fluid.  
This system is of interest in Analogue Gravity~\cite{barcelo2011analogue,barcelo2019analogue}, which aims to mimic field propagation in curved spacetime using laboratory-based systems (such as Bose-Einstein condensates~\cite{garay2000sonic,barcelo2003towards} and nonlinear optics~\cite{Aguero_Santacruz_2020}) and thereby to provide experimental realisations of physical phenomena normally associated with gravity (such as black holes~\cite{braunstein2023analogue} and Hawking radiation~\cite{hawking1974black,hawking1975particle}).  Experiments have been performed with water waves to probe wave scattering at black holes~\cite{Euv__2020} and to observe a classical analogue of the Hawking effect~\cite{Weinfurtner_2011,euve2016observation}.

Reducing the flow to an effective (2+1)-D spacetime requires assumptions about the nature of the flow, in order to remove degrees of freedom associated with the vertical direction.  Typically, we assume that the flow is both incompressible and irrotational.  However, realistic flows tend to exhibit some vorticity due to friction at the bottom of the channel and recirculation after passage over an obstacle.  It thus behooves us to investigate how the presence of vorticity affects the field theory description of waves in this system.
In Analogue Gravity, previous works have explored the effects of vorticity in the context of analogue rotating black holes~\cite{PhysRevD.99.044025,fischer2003space,Perez_Bergliaffa_2004,churilov2019scattering}, but this vorticity is directly inherited by ${\bf u}_{\parallel}$ in~(\ref{eq:PG_line_element}) whereas we are interested here in a ``hidden'' vorticity due to shear flow that has no obvious imprint on the effective metric.
In fluid dynamics, systems that account for vorticity effects have been studied to understand how surface waves are affected~\cite{fabrikant1998propagation}, the simplest case being where a two-dimensional flow has a linear velocity profile~\cite{ellingsen2014linear,maissa2016wave,maissa2016negative}. 
An approach with non-constant vorticity was proposed by Thompson~\cite{thompson1949propagation}, who studied a two-layer flow with non-zero vorticity in the lower layer only and concluded that an additional mode due to the presence of the second layer had emerged.
This is reminiscent of other types of vorticity wave that emerge when the vorticity is oriented longitudinally~\cite{Howe_Liu_1977} or vertically~\cite{VorticityWavesoverStrongTopography}.

In this Letter, we adopt the same flow profile as~\cite{thompson1949propagation}, but we account for the presence of an obstacle to make the flow inhomogeneous.  Unlike~\cite{thompson1949propagation}, we neglect dispersive effects due to finite depth and surface tension (effectively adopting a long-wavelength approximation), but we develop a field theory description that includes two coupled scalar fields and allows us to consider scattering between different modes.

Let us briefly recall the irrotational case~\cite{PhysRevD.66.044019}.
Water flows in an open channel whose bottom is position-dependent.
The flow is assumed to be inviscid, incompressible, and irrotational, and the local flow velocity is thus given by the gradient of a velocity potential: ${\bf u} = -\boldsymbol{\nabla}\varphi$.
We separate the flow into a {\it background} (which is assumed time-independent) and a {\it perturbation}: $\varphi = \varphi_{0} + \delta\varphi$, ${\bf u} = {\bf u}_{0} + \delta{\bf u}$.
For simplicity, we assume that both are independent of the transverse ($y$) direction.
Surface waves are characterised by deformations of the free surface, where the instantaneous water depth $H(t,x)$ is shifted from its background value $h(x)$: $H(t,x) = h(x) + \eta(t,x)$.  For small-amplitude waves ($\left|\eta\right| \ll h$) in the long-wavelength limit ($kh\ll1$ where $k$ is a typical wavenumber and $h$ is the water depth),
it is found that $\delta\varphi$ is independent of $z$. 
Furthermore, $\delta\varphi$ and $\eta$ are found to satisfy the following equations: 
\begin{align}
    \frac{1}{g} \left(\partial_{t} + u_{0} \partial_{x}\right) \delta\varphi &= \eta \,, \nonumber \\
    [(\partial_t + \partial_x u_0)(\partial_t + u_0 \partial_x) - \partial_x c^2 \partial_x]\delta\varphi &= 0 \,,
    \label{eq:eom_no_vorticity}
\end{align}
where $c = \sqrt{g h}$ ($g$ being the acceleration due to gravity) is the local wave speed with respect to the fluid, while $u_{0}$ is the $x$-component of the background flow velocity.
The second of Eqs.~(\ref{eq:eom_no_vorticity}) is precisely the wave equation for a massless scalar field living in a curved spacetime characterised by the metric~(\ref{eq:PG_line_element}) (assuming no $y$-dependence).
The field admits two independent excitations with phase velocities $v_{p} = u_{0} \pm c$; if $u_{0} > 0$, the plus and minus solutions define the {\it co-current} and {\it counter-current} modes, respectively. 
An analogue black hole is realised when the fluid passes from a subcritical region ($u_0 < c$) to a supercritical one ($u_0 > c$). 
In the supercritical region, which is identified with the ``interior'' of the black hole, both co- and counter-current modes are transported in the same direction as the current. 
Moreover, the counter-current mode has negative energy there, in the sense that it acts counter to the background flow and actually reduces the total energy of the system~\cite{LAOstrovski_1986}.  The existence of this negative-energy mode allows {\it anomalous scattering}: an effective extraction of energy from the background flow and a crucial ingredient in many Analogue Gravity contexts, such as superradiance and the Hawking effect.

\begin{figure}
    \includegraphics[width=1\linewidth]{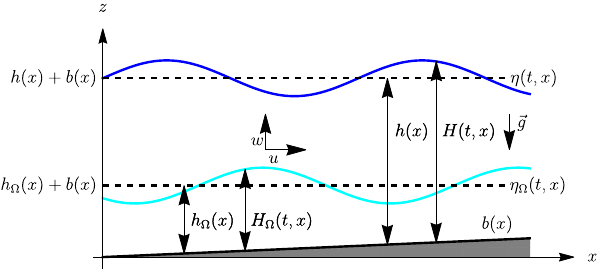}
    \caption{Picture of gravity waves in the channel flow and the relevant parameters. The free surface is placed at $z=b(x)+H(t,x)=b(x)+h(x)+\eta(t,x)$ and the undisturbed surface is at position $z=b(x)+h(x)$. The vorticity is not null for $b(x)\le z \le b(x)+H_\Omega(t,x)$, where $H_\Omega(t,x)=h_\Omega(x)+\eta_\Omega(t,x)$.}
    \label{fig:system}
\end{figure}

Let us now relax the assumption of irrotationality.
Inspired by experimental observations, which indicate that a significant shear is typically generated near the bottom of the channel, we consider the simplified framework shown in Fig.~\ref{fig:system}, where the flow is divided into two layers (upper and lower).
The upper layer is irrotational, while in the lower layer we consider a shear flow that generates a constant non-zero vorticity $\boldsymbol{\Omega}=\boldsymbol{\nabla} \times {\bf u} = \Omega\hat{\textbf{y}}$.
The vorticity $\Omega$ is the only parameter that does not depend on the position $x$, since in a two-dimensional flow the vorticity is constant along streamlines.
Despite this more complicated scenario, we find that, in the limit of small amplitude and long wavelength, the velocity perturbations $\delta u^{\pm}$ (where the $+$ ($-$) superscript refers to the upper (lower) layer) are each $z$-independent.  We may thus introduce two scalar fields, $\delta\varphi^{+}$ and $\delta\varphi^{\Omega}$, such that
\begin{equation}
    \delta u^{+} = -\partial_{x} \delta\varphi^{+} \,, \qquad
    \delta u^{+} - \delta u^{-} = -\partial_{x} \delta\varphi^{\Omega} \,.
\end{equation}
The coupled equations of motion for $\delta\varphi^+$ and $\delta\varphi^{\Omega}$ are (see Appendix~\ref{app:em} for
the derivation)
\begin{align}
    &\left[(\partial_t+\partial_x\unotp)(\partial_t+\unotp\partial_x)-\partial_x gh\partial_x\right]\delta\varphi^+ = -\partial_xgh_\Omega \partial_{x} \delta\varphi^{\Omega} \,, \nonumber \\
    &(\partial_t+(\unotp-\Omega h_\Omega)\partial_x)\delta\varphi^\Omega = -\Omega h_\Omega\partial_x\delta\varphi^+.
    \label{eqs:mot}
\end{align}
The corresponding surface deformations are given by
\begin{equation}
    \eta = \frac{1}{g} \left( \partial_{t} + u^{+}_0 \partial_{x} \right) \delta\varphi^{+} \,, \qquad
    \eta^{\Omega} = -\frac{1}{\Omega} \partial_{x} \delta\varphi^{\Omega} \,.
    \label{eqs:free_surface_deformations}
\end{equation}
The dispersion relation is found by assuming a constant background and that the fields take the form of a plane wave: $\delta\varphi^{+}, \delta\varphi^{\Omega} \propto e^{i k x - i \omega t}$ with $\omega$ and $k$ the frequency and the wave vector, respectively.
Inserting such solutions in Eqs.~\eqref{eqs:mot}, we find 
\begin{multline}
\left(v_{p}-\widetilde{u}_{0}\right)^{3} - \left(g h + \frac{1}{3} \Omega^{2} h_{\Omega}^{2}\right) \left(v_{p}-\widetilde{u}_{0}\right) \\
-2 \Omega h_{\Omega} \left(\frac{1}{3} g h - \frac{1}{2} g h_{\Omega} - \frac{1}{27} \Omega^{2} h_{\Omega}^{2} \right) = 0 \,,
\label{eq:disp_rel}
\end{multline}
where we have substituted the phase velocity $v_{p} = \omega/k$ and defined $\widetilde{u}_{0} = u_{0}^{+} - \frac{1}{3}\Omega h_{\Omega}$. 
Given the cubic nature of Eq.~(\ref{eq:disp_rel}), in general there exist three solutions for $v_{p}$, and as the coefficients entering Eq.~(\ref{eq:disp_rel}) are independent of $\omega$ and $k$, these solutions are non-dispersive.
(See~\cite{thompson1949propagation} for a similar treatment of the dispersive case.) 
The existence of three solutions represents a novelty with respect to the usual situation in Analogue Gravity, where only two surface waves are present.  Indeed, taking $\Omega = 0$ in Eq.~(\ref{eq:disp_rel}) and working in the rest frame of the fluid so that $\widetilde{u}_{0}=0$, we find the solutions $v_{p} = \pm \sqrt{gh}$ and $v_{p} = 0$.  Thus, while mathematically a third solution exists, it is not propagating with respect to the fluid, and the only non-trivial waves are the two surface waves with equal and opposite velocities.  
The presence of vorticity breaks the isotropy of the flow and introduces an additional propagating mode, which we refer to as the \emph{vorticity mode}.  Interestingly, while the presence of this additional mode is related to the necessity of a second scalar field, this additional field generates only one additional mode, rather than two as might have been expected.  This can be traced back to the fact that the second of Eqs.~(\ref{eqs:mot}) is only first-order in time.

Through Eq.~(\ref{eq:disp_rel}), the dependence of the phase velocities of the three modes on the parameters of the flow is rather complicated.  
In App.~\ref{subsec:disp_rel}, we show 
that it is possible to derive certain relations by a careful study of the coefficients of the cubic polynomial.  Here we mention that the phase velocity of the vorticity mode is always between $u_{0}^{+}$ (the flow velocity at the free surface) and $u_{0}^{+}-\Omega h_{\Omega}$ (the flow velocity on the bottom of the channel).  It is thus typically propagating in the direction of the flow, but is also clearly counter-propagating from the point of view of the free surface.  By analogy with the behaviour of the counter-current surface mode, this suggests that the energy of the vorticity mode might be negative.  Proving this requires further calculation, and we will show below that, at least for some flows, it is indeed the case.
Moreover, in App.~\ref{subsec:eff_velocities} we show that we can use the velocities of the two surface modes to define an effective flow velocity $u_{\rm eff}$ and wave speed $c_{\rm eff}$ such that the surface-wave velocities are $u_{\rm eff} \pm c_{\rm eff}$.  These define an effective background, characterising how the flow is affected as far as these ``standard'' modes are concerned.

\begin{figure}
\includegraphics[width=1\linewidth]{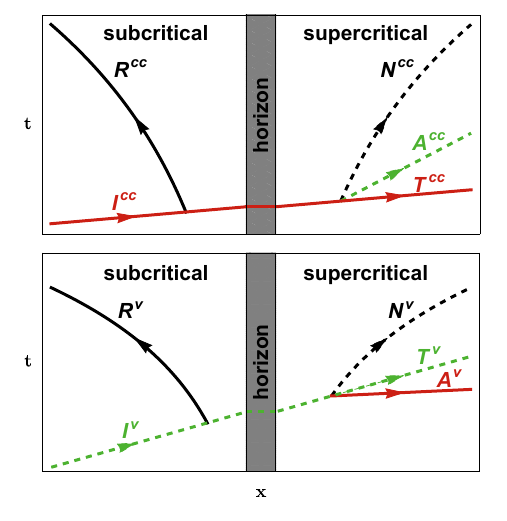}
\caption{Scattering process resulting from sending an incident wave into the subcritical region towards the horizon. Top: the incident wave $I^{\mathrm{cc}}$ is a co-current mode as in \cite{Euv__2020}. Bottom: the incident wave $I^{\mathrm{v}}$ is a vorticity mode. In both images, the colors red, black, and green represent the co-current mode, counter-current mode, and vorticity mode, respectively. The dashed lines are the modes with negative energy.}
\label{fig:scattering}
\end{figure}

Finally, let us turn to an examination of wave scattering at an analogue black hole horizon, in a similar vein to the situation studied in~\cite{Euv__2020}.  The advantage of having derived the relevant wave equations (rather than just the dispersion relation) is that it allows us to calculate how the different modes mix into each other due to inhomogeneities of the background flow.
In the non-dispersive limit, wave scattering occurs~\cite{Euv__2020, euve2017classical,fourdrinoy2022correlations,dewitt1975quantum} as if there were an effective potential~\cite{PhysRevD.87.124018, PhysRevD.90.104044}, and the effective spacetime can be probed by sending incident waves in the subcritical region towards the horizon and observing the products of this process. In the subcritical region, in the laboratory frame, two modes propagate in the same direction as the current. Therefore, there are two possible scenarii: sending an incident wave as a co-current mode or as a vorticity mode. A schematic representation of these scattering processes is shown in Fig.~\ref{fig:scattering}.
The entire process can be split into two sequential scattering events, each confined to one side of the horizon.

Consider first the situation represented in the top panel of Fig.~\ref{fig:scattering}, where an incident co-current wave is incident on the horizon.
It is partially scattered outside the black hole, generating a reflected counter-current wave characterized by the reflection coefficient $R^{\mathrm{cc}}$. As this counter-current wave is in the subcritical region, its energy is positive. 
The unscattered portion of the wave passes through the horizon and scatters inside the black hole, generating a transmitted co-current wave, a transmitted counter-current wave with negative energy (being in the supercritical region), and a 
vorticity wave.  This last one represents an additional product with respect to the standard scattering process studied in~\cite{Euv__2020}.
The waves generated in the supercritical region are characterized by the following amplitudes: the transmission coefficient $T^{\mathrm{cc}}$; the amplitude of the negative-energy counter-current wave, $N^{\mathrm{cc}}$; and the amplitude of the ``additional'' (vorticity) wave, $A^{\mathrm{cc}}$.

When $\Omega \neq 0$, there exists another scattering process where an incident vorticity wave is scattered into the same set of products.  This is schematically represented in the lower panel of Fig.~\ref{fig:scattering}.  Notationally, the only difference with respect to the case above is that the transmission amplitude $T^{\rm v}$ now naturally refers to the transmitted vorticity mode, while the ``additional'' amplitude $A^{\rm v}$ now applies to the co-current mode generated in the supercritical region.

Energy conservation imposes {\it unitarity relations} between these scattering amplitudes.  Since some of the waves carry negative energy -- in particular, the counter-current wave in the supercritical region, and the vorticity wave in both regions -- the corresponding terms in the unitarity relations carry negative signs, and we have
\begin{align}
    1 &= |T^{\mathrm{cc}}|^2+|R^{\mathrm{cc}}|^2-|N^{\mathrm{cc}}|^2-|A^{\mathrm{cc}}|^2 \,, \nonumber \\
    -1 &= -|T^{\mathrm{v}}|^2+|R^{\mathrm{v}}|^2-|N^{\mathrm{v}}|^2+|A^{\mathrm{v}}|^2 \,.
\end{align}
The appearance of both plus and minus signs in the above unitarity relations is indicative of anomalous scattering. 

\begin{figure}
    \includegraphics[width=1.0\linewidth]{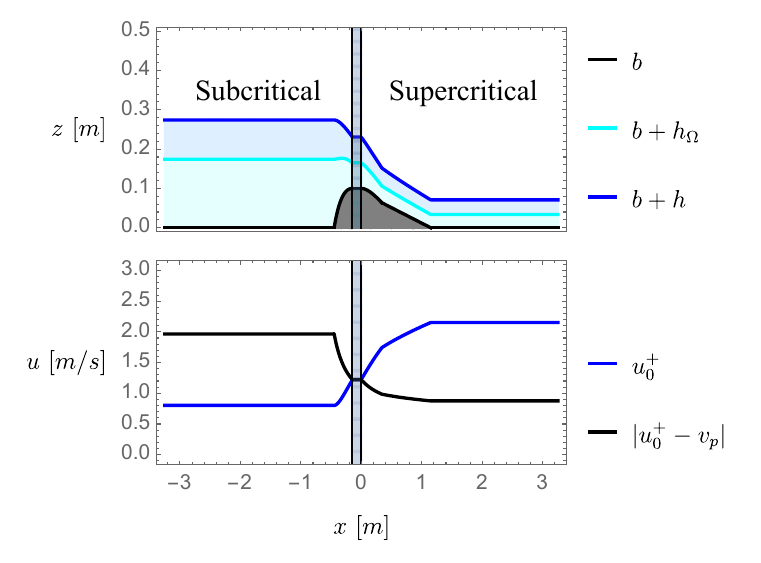}
    \caption{System schemes for which we obtain the scattering coefficients. Top: channel flow profile, the bottom $b(x)$ (black), the interface between the two layers $h_\Omega(x)$ (cyan) and the top surface (blue) are shown. The bottom features are explained in App.~\ref{app:em}. Bottom: velocities profile. The speed of the current on the top $\unotp$ (blue) and the phase velocity for the counter-current mode in the frame which moves with $\unotp$ (black) are shown. For both the pictures, the blurry region represent the points for which the current is transcritical, thus for which there is the analogue horizon. To realise this background we have considered the following parameters: the top layer flux $q_+=0.08\ \mathrm{m}^2/\mathrm{s} $, the bottom layer flux $q_-=0.07\ \mathrm{m}^2/\mathrm{s} $ and the vorticity in the bottom layer $\Omega=4.6\ \mathrm{s}^{-1} $.}
    \label{fig:systemsc}
\end{figure}
\begin{figure}
    \includegraphics[width=1\linewidth]{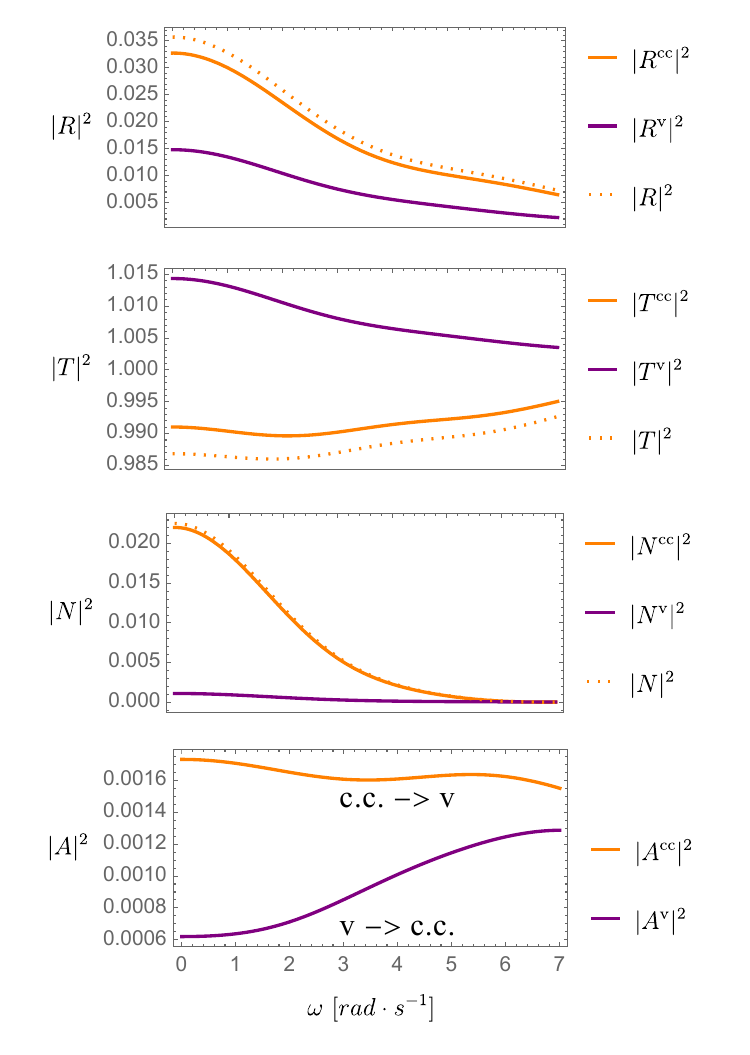}
    \caption{Scattering coefficients for the system shown in Figs.~\ref{fig:systemsc} and for the model with no vorticity with same total flux. The coefficients are displayed for the case where the incident wave is the co-current mode (solid orange), for the case where the incident wave is the vorticity mode (solid purple) and for the case without vorticity (dotted orange).}
    \label{fig:scatteringcoeff}
\end{figure}

In Fig.~\ref{fig:scatteringcoeff} we present the scattering coefficients (for the two scattering processes of Fig.~\ref{fig:scattering}) as functions of the frequency $\omega$.
Details of the numerical method are given in App.~\ref{app:sa}. 
The background used is shown in Fig.~\ref{fig:systemsc}, and has been realised considering the approach explained in App.~\ref{app:em}, 
and we have considered the following parameters: the top layer flux $q_+=0.08\ \mathrm{m}^2/\mathrm{s} $, the bottom layer flux $q_-=0.07\ \mathrm{m}^2/\mathrm{s} $ and the vorticity in the bottom layer $\Omega=4.6\ \mathrm{s}^{-1}$. We have chosen to fix these parameters because they are the only quantities that are $x$-independent and are thus conserved in the system.
The scattering coefficients associated to an incident co-current (vorticity) mode are shown in orange (purple), while in dotted orange are shown those coefficients for a flow with no vorticity but the same total flux.  (Note that the no-vorticity case only applies to the incident co-current mode, and produces no outgoing vorticity mode.)
This indicates that the presence of vorticity affects mainly the reflection and transmission coefficients, with $\left|R\right|^{2}$ being reduced by ~10\%.
(In App.~\ref{app:sa}, 
we show the percentage difference in the scattering coefficients with respect to the case without vorticity, for different values of $\Omega$ and always with the same total flux.)
The scattering coefficients associated with an incident vorticity mode 
are seen to be typically much smaller than the other scattering coefficients, with most of the wave being
either reflected or transmitted. Importantly, the vorticity wave's transmission coefficient is greater than 1; this confirms that the vorticity wave and the reflected counter-current wave have opposite energies, and thus that the vorticity wave has negative energy. 

To summarize, we have generalized the standard field theory treatment of surface waves in a 1D channel flow to incorporate the presence of a bottom layer of vorticity due to a shear flow in the vertical direction.  
The model introduces a second scalar field that couples to the usual velocity potential.
We have derived wave equations for these fields and found the dispersion relation associated to its normal modes, which demonstrate that the presence of vorticity can influence the behaviour of surface waves and introduces another mode 
characterised by negative energy and predominantly localised at the interface between layers with and without vorticity.
Using this model, we were able to calculate the scattering coefficients during wave scattering events. Our results show that, given a fixed bottom shape and total flux, the scattering coefficients are affected by the presence of vorticity, depending on the flux in each layer and the vorticity itself. Specifically, when the bottom layer is much smaller than the top layer ($h_\Omega/h \ll 1$), the scattering coefficients obtained from the model with vorticity (when the co-current mode is the incident wave) are almost identical to those obtained from the model without vorticity, with discrepancies less than $1\%$. This confirms the robustness of the results found in~\cite{Euv__2020}. However, when the bottom layer is much larger than the top layer ($h_\Omega/h \sim 1$), the scattering coefficients from the model with vorticity (for the co-current mode) differ significantly from those of the model without vorticity, showing discrepancies greater than $10\%$ (see App.~\ref{app:sa}).

The implications of our work are multifaceted. The inclusion of vorticity, which in practice is always present to some extent, 
not only enhances the accuracy of our theoretical predictions 
but also allows us to understand the limits within which vorticity can be neglected.
Within the context of Analogue Gravity, it can be viewed as a study into the possible effects of an internal structure of spacetime, here provided by the non-trivial velocity profile in the vertical direction that does not enter explicitly in the effective metric seen by surface waves.
We hope that future experiments will be able to 
test the predictions of our vorticity-inclusive model. 
On the theoretical side, 
extending these results to the dispersive case~\cite{fabrikant1998propagation,coutant2014hawking} could provide deeper insights into the underlying physics and broaden the applicability of our findings,
particularly to two-obstacle configurations designed to probe the black hole laser effect~\cite{BHLpreprint}.
Additionally, we aim to extend this model to three dimensions and to determine whether it is possible to define an effective analogue metric and to explore the role of vorticity in the context of Analogue Gravity. The possibility of an internal structure in gravitational spacetimes remains, of course, speculative.

\section*{Acknowledgments}
AB and SR are funded by the CNRS Chair in Physical Hydrodynamics (reference ANR-22-CPJ2-0039-01).
This work pertains (namely, is not funded but enters in the scientific perimeter) to the French government programs ``Investissement d'Avenir'' EUR INTREE (reference ANR-18-EURE-0010) and LABEX INTERACTIFS (reference ANR-11-LABX-0017-01).

\bibliography{bib}

\begin{thebibliography}{39}%
\makeatletter
\providecommand \@ifxundefined [1]{%
 \@ifx{#1\undefined}
}%
\providecommand \@ifnum [1]{%
 \ifnum #1\expandafter \@firstoftwo
 \else \expandafter \@secondoftwo
 \fi
}%
\providecommand \@ifx [1]{%
 \ifx #1\expandafter \@firstoftwo
 \else \expandafter \@secondoftwo
 \fi
}%
\providecommand \natexlab [1]{#1}%
\providecommand \enquote  [1]{``#1''}%
\providecommand \bibnamefont  [1]{#1}%
\providecommand \bibfnamefont [1]{#1}%
\providecommand \citenamefont [1]{#1}%
\providecommand \href@noop [0]{\@secondoftwo}%
\providecommand \href [0]{\begingroup \@sanitize@url \@href}%
\providecommand \@href[1]{\@@startlink{#1}\@@href}%
\providecommand \@@href[1]{\endgroup#1\@@endlink}%
\providecommand \@sanitize@url [0]{\catcode `\\12\catcode `\$12\catcode
  `\&12\catcode `\#12\catcode `\^12\catcode `\_12\catcode `\%12\relax}%
\providecommand \@@startlink[1]{}%
\providecommand \@@endlink[0]{}%
\providecommand \url  [0]{\begingroup\@sanitize@url \@url }%
\providecommand \@url [1]{\endgroup\@href {#1}{\urlprefix }}%
\providecommand \urlprefix  [0]{URL }%
\providecommand \Eprint [0]{\href }%
\providecommand \doibase [0]{http://dx.doi.org/}%
\providecommand \selectlanguage [0]{\@gobble}%
\providecommand \bibinfo  [0]{\@secondoftwo}%
\providecommand \bibfield  [0]{\@secondoftwo}%
\providecommand \translation [1]{[#1]}%
\providecommand \BibitemOpen [0]{}%
\providecommand \bibitemStop [0]{}%
\providecommand \bibitemNoStop [0]{.\EOS\space}%
\providecommand \EOS [0]{\spacefactor3000\relax}%
\providecommand \BibitemShut  [1]{\csname bibitem#1\endcsname}%
\let\auto@bib@innerbib\@empty
\bibitem [{\citenamefont {Sch\"utzhold}\ and\ \citenamefont
  {Unruh}(2002)}]{PhysRevD.66.044019}%
  \BibitemOpen
  \bibfield  {author} {\bibinfo {author} {\bibfnamefont {Ralf}\ \bibnamefont
  {Sch\"utzhold}}\ and\ \bibinfo {author} {\bibfnamefont {William~G.}\
  \bibnamefont {Unruh}},\ }\bibfield  {title} {\enquote {\bibinfo {title}
  {Gravity wave analogues of black holes},}\ }\href {\doibase
  10.1103/PhysRevD.66.044019} {\bibfield  {journal} {\bibinfo  {journal} {Phys.
  Rev. D}\ }\textbf {\bibinfo {volume} {66}},\ \bibinfo {pages} {044019}
  (\bibinfo {year} {2002})}\BibitemShut {NoStop}%
\bibitem [{\citenamefont {Rousseaux}(2013)}]{rousseaux2013basics}%
  \BibitemOpen
  \bibfield  {author} {\bibinfo {author} {\bibfnamefont {Germain}\ \bibnamefont
  {Rousseaux}},\ }\enquote {\bibinfo {title} {The basics of water waves theory
  for analogue gravity},}\ in\ \href {\doibase 10.1007/978-3-319-00266-8_5}
  {\emph {\bibinfo {booktitle} {Analogue Gravity Phenomenology: Analogue
  Spacetimes and Horizons, from Theory to Experiment}}},\ \bibinfo {editor}
  {edited by\ \bibinfo {editor} {\bibfnamefont {Daniele}\ \bibnamefont
  {Faccio}}, \bibinfo {editor} {\bibfnamefont {Francesco}\ \bibnamefont
  {Belgiorno}}, \bibinfo {editor} {\bibfnamefont {Sergio}\ \bibnamefont
  {Cacciatori}}, \bibinfo {editor} {\bibfnamefont {Vittorio}\ \bibnamefont
  {Gorini}}, \bibinfo {editor} {\bibfnamefont {Stefano}\ \bibnamefont
  {Liberati}}, \ and\ \bibinfo {editor} {\bibfnamefont {Ugo}\ \bibnamefont
  {Moschella}}}\ (\bibinfo  {publisher} {Springer International Publishing},\
  \bibinfo {address} {Cham},\ \bibinfo {year} {2013})\ pp.\ \bibinfo {pages}
  {81--107}\BibitemShut {NoStop}%
\bibitem [{\citenamefont {Painlev{\'e}}(1921)}]{painleve1921mecanique}%
  \BibitemOpen
  \bibfield  {author} {\bibinfo {author} {\bibfnamefont {Paul}\ \bibnamefont
  {Painlev{\'e}}},\ }\bibfield  {title} {\enquote {\bibinfo {title} {La
  m{\'e}canique classique et la th{\'e}orie de la relativit{\'e}},}\
  }\href@noop {} {\bibfield  {journal} {\bibinfo  {journal} {Comptes Rendus de
  l'Académie des Sciences (série non spécifiée)}\ }\textbf {\bibinfo
  {volume} {173}},\ \bibinfo {pages} {677--680} (\bibinfo {year}
  {1921})}\BibitemShut {NoStop}%
\bibitem [{\citenamefont {Barcelo}\ \emph {et~al.}(2011)\citenamefont
  {Barcelo}, \citenamefont {Liberati},\ and\ \citenamefont
  {Visser}}]{barcelo2011analogue}%
  \BibitemOpen
  \bibfield  {author} {\bibinfo {author} {\bibfnamefont {C.}~\bibnamefont
  {Barcelo}}, \bibinfo {author} {\bibfnamefont {S.}~\bibnamefont {Liberati}}, \
  and\ \bibinfo {author} {\bibfnamefont {M.}~\bibnamefont {Visser}},\
  }\bibfield  {title} {\enquote {\bibinfo {title} {Analogue gravity},}\ }\href
  {https://link.springer.com/article/10.12942/lrr-2011-3} {\bibfield  {journal}
  {\bibinfo  {journal} {Living Rev. Relativ.}\ }\textbf {\bibinfo {volume}
  {14}},\ \bibinfo {pages} {1--159} (\bibinfo {year} {2011})}\BibitemShut
  {NoStop}%
\bibitem [{\citenamefont {Barceló}(2019)}]{barcelo2019analogue}%
  \BibitemOpen
  \bibfield  {author} {\bibinfo {author} {\bibfnamefont {Carlos}\ \bibnamefont
  {Barceló}},\ }\bibfield  {title} {\enquote {\bibinfo {title} {Analogue
  black-hole horizons},}\ }\href {\doibase 10.1038/s41567-018-0367-6}
  {\bibfield  {journal} {\bibinfo  {journal} {Nature Physics}\ }\textbf
  {\bibinfo {volume} {15}},\ \bibinfo {pages} {210--213} (\bibinfo {year}
  {2019})}\BibitemShut {NoStop}%
\bibitem [{\citenamefont {Garay}\ \emph {et~al.}(2000)\citenamefont {Garay},
  \citenamefont {Anglin}, \citenamefont {Cirac},\ and\ \citenamefont
  {Zoller}}]{garay2000sonic}%
  \BibitemOpen
  \bibfield  {author} {\bibinfo {author} {\bibfnamefont {L.~J.}\ \bibnamefont
  {Garay}}, \bibinfo {author} {\bibfnamefont {J.~R.}\ \bibnamefont {Anglin}},
  \bibinfo {author} {\bibfnamefont {J.~I.}\ \bibnamefont {Cirac}}, \ and\
  \bibinfo {author} {\bibfnamefont {P.}~\bibnamefont {Zoller}},\ }\bibfield
  {title} {\enquote {\bibinfo {title} {Sonic analog of gravitational black
  holes in {Bose-Einstein Condensates}},}\ }\href {\doibase
  10.1103/PhysRevLett.85.4643} {\bibfield  {journal} {\bibinfo  {journal}
  {Phys. Rev. Lett.}\ }\textbf {\bibinfo {volume} {85}},\ \bibinfo {pages}
  {4643--4647} (\bibinfo {year} {2000})}\BibitemShut {NoStop}%
\bibitem [{\citenamefont {Barcelo}\ \emph {et~al.}(2003)\citenamefont
  {Barcelo}, \citenamefont {Liberati},\ and\ \citenamefont
  {Visser}}]{barcelo2003towards}%
  \BibitemOpen
  \bibfield  {author} {\bibinfo {author} {\bibfnamefont {Carlos}\ \bibnamefont
  {Barcelo}}, \bibinfo {author} {\bibfnamefont {Stefano}\ \bibnamefont
  {Liberati}}, \ and\ \bibinfo {author} {\bibfnamefont {Matt}\ \bibnamefont
  {Visser}},\ }\bibfield  {title} {\enquote {\bibinfo {title} {Towards the
  observation of {Hawking} radiation in {Bose--Einstein} condensates},}\ }\href
  {https://www.worldscientific.com/doi/abs/10.1142/S0217751X0301615X?casa_token=QLARLJ4jXNUAAAAA%3AKbPyidcwTtuvhCrmfa6yVn7_gGESlzafKVLmayBOhXXZjAHyzAL_9HK7p3abXM7W5NqGgUBVgF5P5A}
  {\bibfield  {journal} {\bibinfo  {journal} {Int. J. Mod. Phys. A}\ }\textbf
  {\bibinfo {volume} {18}},\ \bibinfo {pages} {3735--3745} (\bibinfo {year}
  {2003})}\BibitemShut {NoStop}%
\bibitem [{\citenamefont {Aguero-Santacruz}\ and\ \citenamefont
  {Bermudez}(2020)}]{Aguero_Santacruz_2020}%
  \BibitemOpen
  \bibfield  {author} {\bibinfo {author} {\bibfnamefont {Raul}\ \bibnamefont
  {Aguero-Santacruz}}\ and\ \bibinfo {author} {\bibfnamefont {David}\
  \bibnamefont {Bermudez}},\ }\bibfield  {title} {\enquote {\bibinfo {title}
  {Hawking radiation in optics and beyond},}\ }\href
  {https://royalsocietypublishing.org/doi/10.1098/rsta.2019.0223} {\bibfield
  {journal} {\bibinfo  {journal} {Philosophical Transactions of the Royal
  Society A}\ }\textbf {\bibinfo {volume} {378}},\ \bibinfo {pages} {20190223}
  (\bibinfo {year} {2020})}\BibitemShut {NoStop}%
\bibitem [{\citenamefont {Braunstein}\ \emph {et~al.}(2023)\citenamefont
  {Braunstein}, \citenamefont {Faizal}, \citenamefont {Krauss}, \citenamefont
  {Marino},\ and\ \citenamefont {Shah}}]{braunstein2023analogue}%
  \BibitemOpen
  \bibfield  {author} {\bibinfo {author} {\bibfnamefont {Samuel~L.}\
  \bibnamefont {Braunstein}}, \bibinfo {author} {\bibfnamefont {Mir}\
  \bibnamefont {Faizal}}, \bibinfo {author} {\bibfnamefont {Lawrence~M.}\
  \bibnamefont {Krauss}}, \bibinfo {author} {\bibfnamefont {Francesco}\
  \bibnamefont {Marino}}, \ and\ \bibinfo {author} {\bibfnamefont {Naveed~A.}\
  \bibnamefont {Shah}},\ }\bibfield  {title} {\enquote {\bibinfo {title}
  {Analogue simulations of quantum gravity with fluids},}\ }\href {\doibase
  10.1038/s42254-023-00630-y} {\bibfield  {journal} {\bibinfo  {journal}
  {Nature Reviews Physics}\ }\textbf {\bibinfo {volume} {5}},\ \bibinfo {pages}
  {612--622} (\bibinfo {year} {2023})}\BibitemShut {NoStop}%
\bibitem [{\citenamefont {Hawking}(1974)}]{hawking1974black}%
  \BibitemOpen
  \bibfield  {author} {\bibinfo {author} {\bibfnamefont {S.~W.}\ \bibnamefont
  {Hawking}},\ }\bibfield  {title} {\enquote {\bibinfo {title} {Black hole
  explosions?}}\ }\href {\doibase 10.1038/248030a0} {\bibfield  {journal}
  {\bibinfo  {journal} {Nature}\ }\textbf {\bibinfo {volume} {248}},\ \bibinfo
  {pages} {30--31} (\bibinfo {year} {1974})}\BibitemShut {NoStop}%
\bibitem [{\citenamefont {Hawking}(1975)}]{hawking1975particle}%
  \BibitemOpen
  \bibfield  {author} {\bibinfo {author} {\bibfnamefont {Stephen~W}\
  \bibnamefont {Hawking}},\ }\bibfield  {title} {\enquote {\bibinfo {title}
  {Particle creation by black holes},}\ }\href
  {https://idp.springer.com/authorize/casa?redirect_uri=https://link.springer.com/article/10.1007/BF02345020&casa_token=X9Qt-MqI3WQAAAAA:-WutRYFk8nV4rZZj6pcuUn7mQpPKWVETmsjgjxF2Pz4lira7b0mDy9WxUUEba_VomNlm_k8fZFUt_N2hcS0}
  {\bibfield  {journal} {\bibinfo  {journal} {Comm. Math. Phys.}\ }\textbf
  {\bibinfo {volume} {43}},\ \bibinfo {pages} {199--220} (\bibinfo {year}
  {1975})}\BibitemShut {NoStop}%
\bibitem [{\citenamefont {Euv\'e}\ \emph {et~al.}(2020)\citenamefont {Euv\'e},
  \citenamefont {Robertson}, \citenamefont {James}, \citenamefont {Fabbri},\
  and\ \citenamefont {Rousseaux}}]{Euv__2020}%
  \BibitemOpen
  \bibfield  {author} {\bibinfo {author} {\bibfnamefont {L\'eo-Paul}\
  \bibnamefont {Euv\'e}}, \bibinfo {author} {\bibfnamefont {Scott}\
  \bibnamefont {Robertson}}, \bibinfo {author} {\bibfnamefont {Nicolas}\
  \bibnamefont {James}}, \bibinfo {author} {\bibfnamefont {Alessandro}\
  \bibnamefont {Fabbri}}, \ and\ \bibinfo {author} {\bibfnamefont {Germain}\
  \bibnamefont {Rousseaux}},\ }\bibfield  {title} {\enquote {\bibinfo {title}
  {Scattering of co-current surface waves on an analogue black hole},}\ }\href
  {\doibase 10.1103/PhysRevLett.124.141101} {\bibfield  {journal} {\bibinfo
  {journal} {Phys. Rev. Lett.}\ }\textbf {\bibinfo {volume} {124}},\ \bibinfo
  {pages} {141101} (\bibinfo {year} {2020})}\BibitemShut {NoStop}%
\bibitem [{\citenamefont {Weinfurtner}\ \emph {et~al.}(2011)\citenamefont
  {Weinfurtner}, \citenamefont {Tedford}, \citenamefont {Penrice},
  \citenamefont {Unruh},\ and\ \citenamefont {Lawrence}}]{Weinfurtner_2011}%
  \BibitemOpen
  \bibfield  {author} {\bibinfo {author} {\bibfnamefont {Silke}\ \bibnamefont
  {Weinfurtner}}, \bibinfo {author} {\bibfnamefont {Edmund~W.}\ \bibnamefont
  {Tedford}}, \bibinfo {author} {\bibfnamefont {Matthew C.~J.}\ \bibnamefont
  {Penrice}}, \bibinfo {author} {\bibfnamefont {William~G.}\ \bibnamefont
  {Unruh}}, \ and\ \bibinfo {author} {\bibfnamefont {Gregory~A.}\ \bibnamefont
  {Lawrence}},\ }\bibfield  {title} {\enquote {\bibinfo {title} {Measurement of
  stimulated {Hawking} emission in an analogue system},}\ }\href {\doibase
  10.1103/PhysRevLett.106.021302} {\bibfield  {journal} {\bibinfo  {journal}
  {Phys. Rev. Lett.}\ }\textbf {\bibinfo {volume} {106}},\ \bibinfo {pages}
  {021302} (\bibinfo {year} {2011})}\BibitemShut {NoStop}%
\bibitem [{\citenamefont {Euv\'e}\ \emph {et~al.}(2016)\citenamefont {Euv\'e},
  \citenamefont {Michel}, \citenamefont {Parentani}, \citenamefont {Philbin},\
  and\ \citenamefont {Rousseaux}}]{euve2016observation}%
  \BibitemOpen
  \bibfield  {author} {\bibinfo {author} {\bibfnamefont {L.-P.}\ \bibnamefont
  {Euv\'e}}, \bibinfo {author} {\bibfnamefont {F.}~\bibnamefont {Michel}},
  \bibinfo {author} {\bibfnamefont {R.}~\bibnamefont {Parentani}}, \bibinfo
  {author} {\bibfnamefont {T.~G.}\ \bibnamefont {Philbin}}, \ and\ \bibinfo
  {author} {\bibfnamefont {G.}~\bibnamefont {Rousseaux}},\ }\bibfield  {title}
  {\enquote {\bibinfo {title} {Observation of noise correlated by the {Hawking}
  effect in a water tank},}\ }\href {\doibase 10.1103/PhysRevLett.117.121301}
  {\bibfield  {journal} {\bibinfo  {journal} {Phys. Rev. Lett.}\ }\textbf
  {\bibinfo {volume} {117}},\ \bibinfo {pages} {121301} (\bibinfo {year}
  {2016})}\BibitemShut {NoStop}%
\bibitem [{\citenamefont {Liberati}\ \emph {et~al.}(2019)\citenamefont
  {Liberati}, \citenamefont {Schuster}, \citenamefont {Tricella},\ and\
  \citenamefont {Visser}}]{PhysRevD.99.044025}%
  \BibitemOpen
  \bibfield  {author} {\bibinfo {author} {\bibfnamefont {Stefano}\ \bibnamefont
  {Liberati}}, \bibinfo {author} {\bibfnamefont {Sebastian}\ \bibnamefont
  {Schuster}}, \bibinfo {author} {\bibfnamefont {Giovanni}\ \bibnamefont
  {Tricella}}, \ and\ \bibinfo {author} {\bibfnamefont {Matt}\ \bibnamefont
  {Visser}},\ }\bibfield  {title} {\enquote {\bibinfo {title} {Vorticity in
  analogue spacetimes},}\ }\href {\doibase 10.1103/PhysRevD.99.044025}
  {\bibfield  {journal} {\bibinfo  {journal} {Phys. Rev. D}\ }\textbf {\bibinfo
  {volume} {99}},\ \bibinfo {pages} {044025} (\bibinfo {year}
  {2019})}\BibitemShut {NoStop}%
\bibitem [{\citenamefont {Fischer}\ and\ \citenamefont
  {Visser}(2003)}]{fischer2003space}%
  \BibitemOpen
  \bibfield  {author} {\bibinfo {author} {\bibfnamefont {Uwe~R.}\ \bibnamefont
  {Fischer}}\ and\ \bibinfo {author} {\bibfnamefont {Matt}\ \bibnamefont
  {Visser}},\ }\bibfield  {title} {\enquote {\bibinfo {title} {On the
  space-time curvature experienced by quasiparticle excitations in the
  {Painlevé–Gullstrand} effective geometry},}\ }\href {\doibase
  https://doi.org/10.1016/S0003-4916(03)00011-3} {\bibfield  {journal}
  {\bibinfo  {journal} {Annals of Physics}\ }\textbf {\bibinfo {volume}
  {304}},\ \bibinfo {pages} {22--39} (\bibinfo {year} {2003})}\BibitemShut
  {NoStop}%
\bibitem [{\citenamefont {Perez~Bergliaffa}\ \emph {et~al.}(2004)\citenamefont
  {Perez~Bergliaffa}, \citenamefont {Hibberd}, \citenamefont {Stone},\ and\
  \citenamefont {Visser}}]{Perez_Bergliaffa_2004}%
  \BibitemOpen
  \bibfield  {author} {\bibinfo {author} {\bibfnamefont {Santiago~Esteban}\
  \bibnamefont {Perez~Bergliaffa}}, \bibinfo {author} {\bibfnamefont {Katrina}\
  \bibnamefont {Hibberd}}, \bibinfo {author} {\bibfnamefont {Michael}\
  \bibnamefont {Stone}}, \ and\ \bibinfo {author} {\bibfnamefont {Matt}\
  \bibnamefont {Visser}},\ }\bibfield  {title} {\enquote {\bibinfo {title}
  {Wave equation for sound in fluids with vorticity},}\ }\href {\doibase
  10.1016/j.physd.2003.11.007} {\bibfield  {journal} {\bibinfo  {journal}
  {Physica D: Nonlinear Phenomena}\ }\textbf {\bibinfo {volume} {191}},\
  \bibinfo {pages} {121–136} (\bibinfo {year} {2004})}\BibitemShut {NoStop}%
\bibitem [{\citenamefont {Churilov}\ and\ \citenamefont
  {Stepanyants}(2019)}]{churilov2019scattering}%
  \BibitemOpen
  \bibfield  {author} {\bibinfo {author} {\bibfnamefont {Semyon}\ \bibnamefont
  {Churilov}}\ and\ \bibinfo {author} {\bibfnamefont {Yury}\ \bibnamefont
  {Stepanyants}},\ }\bibfield  {title} {\enquote {\bibinfo {title} {Scattering
  of surface shallow water waves on a draining bathtub vortex},}\ }\href
  {\doibase 10.1103/PhysRevFluids.4.034704} {\bibfield  {journal} {\bibinfo
  {journal} {Phys. Rev. Fluids}\ }\textbf {\bibinfo {volume} {4}},\ \bibinfo
  {pages} {034704} (\bibinfo {year} {2019})}\BibitemShut {NoStop}%
\bibitem [{\citenamefont {Fabrikant}\ and\ \citenamefont
  {Stepanyants}(1998)}]{fabrikant1998propagation}%
  \BibitemOpen
  \bibfield  {author} {\bibinfo {author} {\bibfnamefont {A~L}\ \bibnamefont
  {Fabrikant}}\ and\ \bibinfo {author} {\bibfnamefont {Yu~A}\ \bibnamefont
  {Stepanyants}},\ }\href {\doibase 10.1142/2557} {\emph {\bibinfo {title}
  {Propagation of Waves in Shear Flows}}}\ (\bibinfo  {publisher} {World
  Scientific},\ \bibinfo {year} {1998})\ \Eprint
  {http://arxiv.org/abs/https://www.worldscientific.com/doi/pdf/10.1142/2557}
  {https://www.worldscientific.com/doi/pdf/10.1142/2557} \BibitemShut {NoStop}%
\bibitem [{\citenamefont {Ellingsen}\ and\ \citenamefont
  {Brevik}(2014)}]{ellingsen2014linear}%
  \BibitemOpen
  \bibfield  {author} {\bibinfo {author} {\bibfnamefont {Simen~{\AA}}\
  \bibnamefont {Ellingsen}}\ and\ \bibinfo {author} {\bibfnamefont {Iver}\
  \bibnamefont {Brevik}},\ }\bibfield  {title} {\enquote {\bibinfo {title} {How
  linear surface waves are affected by a current with constant vorticity},}\
  }\href
  {https://iopscience.iop.org/article/10.1088/0143-0807/35/2/025005/meta?casa_token=fUE7nR1FSM0AAAAA:4DyrSC7DSItW_sE6wPOfTqhINgWorM6x8nAOatCCW0FxDN0t9gy8nPIB0ZeGDpmIbXiCCd3Eh3PPtB9IZHOsXiP-cBNUqA}
  {\bibfield  {journal} {\bibinfo  {journal} {Eur. J. Phys.}\ }\textbf
  {\bibinfo {volume} {35}},\ \bibinfo {pages} {025005} (\bibinfo {year}
  {2014})}\BibitemShut {NoStop}%
\bibitem [{\citenamefont {Ma{\"\i}ssa}\ \emph
  {et~al.}(2016{\natexlab{a}})\citenamefont {Ma{\"\i}ssa}, \citenamefont
  {Rousseaux},\ and\ \citenamefont {Stepanyants}}]{maissa2016wave}%
  \BibitemOpen
  \bibfield  {author} {\bibinfo {author} {\bibfnamefont {Philippe}\
  \bibnamefont {Ma{\"\i}ssa}}, \bibinfo {author} {\bibfnamefont {Germain}\
  \bibnamefont {Rousseaux}}, \ and\ \bibinfo {author} {\bibfnamefont {Yury}\
  \bibnamefont {Stepanyants}},\ }\bibfield  {title} {\enquote {\bibinfo {title}
  {Wave blocking phenomenon of surface waves on a shear flow with a constant
  vorticity},}\ }\href
  {https://pubs.aip.org/aip/pof/article-abstract/28/3/032102/316280/Wave-blocking-phenomenon-of-surface-waves-on-a?redirectedFrom=fulltext}
  {\bibfield  {journal} {\bibinfo  {journal} {Phys. Fluids}\ }\textbf {\bibinfo
  {volume} {28}},\ \bibinfo {pages} {032102} (\bibinfo {year}
  {2016}{\natexlab{a}})}\BibitemShut {NoStop}%
\bibitem [{\citenamefont {Ma{\"\i}ssa}\ \emph
  {et~al.}(2016{\natexlab{b}})\citenamefont {Ma{\"\i}ssa}, \citenamefont
  {Rousseaux},\ and\ \citenamefont {Stepanyants}}]{maissa2016negative}%
  \BibitemOpen
  \bibfield  {author} {\bibinfo {author} {\bibfnamefont {Philippe}\
  \bibnamefont {Ma{\"\i}ssa}}, \bibinfo {author} {\bibfnamefont {Germain}\
  \bibnamefont {Rousseaux}}, \ and\ \bibinfo {author} {\bibfnamefont {Yury}\
  \bibnamefont {Stepanyants}},\ }\bibfield  {title} {\enquote {\bibinfo {title}
  {Negative energy waves in a shear flow with a linear profile},}\ }\href
  {https://www.sciencedirect.com/science/article/pii/S0997754615300686?casa_token=l8CIiDjHNU8AAAAA:thBgnHLbw_C71h0l-M5gKUJnnZxIMyxTv_lv_HKJfHUaaymAUb0-MxPsSb7qKMjwY9VMtYsjYm3F}
  {\bibfield  {journal} {\bibinfo  {journal} {Eur. J. Mech. B (Fluids)}\
  }\textbf {\bibinfo {volume} {56}},\ \bibinfo {pages} {192--199} (\bibinfo
  {year} {2016}{\natexlab{b}})}\BibitemShut {NoStop}%
\bibitem [{\citenamefont {Thompson}(1949)}]{thompson1949propagation}%
  \BibitemOpen
  \bibfield  {author} {\bibinfo {author} {\bibfnamefont {Philip~Duncan}\
  \bibnamefont {Thompson}},\ }\bibfield  {title} {\enquote {\bibinfo {title}
  {The propagation of small surface disturbances through rotational flow},}\
  }\href
  {https://nyaspubs.onlinelibrary.wiley.com/doi/abs/10.1111/j.1749-6632.1949.tb27285.x}
  {\bibfield  {journal} {\bibinfo  {journal} {Annals of the New York Academy of
  Sciences}\ }\textbf {\bibinfo {volume} {51}},\ \bibinfo {pages} {463--474}
  (\bibinfo {year} {1949})}\BibitemShut {NoStop}%
\bibitem [{\citenamefont {Howe}\ and\ \citenamefont
  {Liu}(1977)}]{Howe_Liu_1977}%
  \BibitemOpen
  \bibfield  {author} {\bibinfo {author} {\bibfnamefont {M.~S.}\ \bibnamefont
  {Howe}}\ and\ \bibinfo {author} {\bibfnamefont {J.~T.~C.}\ \bibnamefont
  {Liu}},\ }\bibfield  {title} {\enquote {\bibinfo {title} {The generation of
  sound by vorticity waves in swirling duct flows},}\ }\href {\doibase
  10.1017/S0022112077002109} {\bibfield  {journal} {\bibinfo  {journal} {J.
  Fluid Mech.}\ }\textbf {\bibinfo {volume} {81}},\ \bibinfo {pages}
  {369–383} (\bibinfo {year} {1977})}\BibitemShut {NoStop}%
\bibitem [{\citenamefont {Gratton}\ and\ \citenamefont
  {LeBlond}(1986)}]{VorticityWavesoverStrongTopography}%
  \BibitemOpen
  \bibfield  {author} {\bibinfo {author} {\bibfnamefont {Yves}\ \bibnamefont
  {Gratton}}\ and\ \bibinfo {author} {\bibfnamefont {Paul~H.}\ \bibnamefont
  {LeBlond}},\ }\bibfield  {title} {\enquote {\bibinfo {title} {Vorticity waves
  over strong topography},}\ }\href {\doibase
  10.1175/1520-0485(1986)016<0151:VWOST>2.0.CO;2} {\bibfield  {journal}
  {\bibinfo  {journal} {J. Phys. Oceanogr.}\ }\textbf {\bibinfo {volume}
  {16}},\ \bibinfo {pages} {151--166} (\bibinfo {year} {1986})}\BibitemShut
  {NoStop}%
\bibitem [{\citenamefont {Ostrovski{\u\i}}\ \emph {et~al.}(1986)\citenamefont
  {Ostrovski{\u\i}}, \citenamefont {Rybak},\ and\ \citenamefont
  {Tsimring}}]{LAOstrovski_1986}%
  \BibitemOpen
  \bibfield  {author} {\bibinfo {author} {\bibfnamefont {LA}~\bibnamefont
  {Ostrovski{\u\i}}}, \bibinfo {author} {\bibfnamefont {Samuil~A}\ \bibnamefont
  {Rybak}}, \ and\ \bibinfo {author} {\bibfnamefont {L~Sh}\ \bibnamefont
  {Tsimring}},\ }\bibfield  {title} {\enquote {\bibinfo {title} {Negative
  energy waves in hydrodynamics},}\ }\href
  {https://iopscience.iop.org/article/10.1070/PU1986v029n11ABEH003538/meta?casa_token=q_wjRCFY7lQAAAAA:DQTOzXY7b8vW2FcREovzXOPvYGh8OwpA_ICSQchpeg3qaW9TnWkSF5F5bHb1cWIXpyM1RH5u4_dM78GZAx_FiKP1vDHdJQ}
  {\bibfield  {journal} {\bibinfo  {journal} {Sov. Phys. Usp.}\ }\textbf
  {\bibinfo {volume} {29}},\ \bibinfo {pages} {1040} (\bibinfo {year}
  {1986})}\BibitemShut {NoStop}%
\bibitem [{\citenamefont {Euv\'e}\ and\ \citenamefont
  {Rousseaux}(2017)}]{euve2017classical}%
  \BibitemOpen
  \bibfield  {author} {\bibinfo {author} {\bibfnamefont {L.-P.}\ \bibnamefont
  {Euv\'e}}\ and\ \bibinfo {author} {\bibfnamefont {G.}~\bibnamefont
  {Rousseaux}},\ }\bibfield  {title} {\enquote {\bibinfo {title} {Classical
  analogue of an interstellar travel through a hydrodynamic wormhole},}\ }\href
  {\doibase 10.1103/PhysRevD.96.064042} {\bibfield  {journal} {\bibinfo
  {journal} {Phys. Rev. D}\ }\textbf {\bibinfo {volume} {96}},\ \bibinfo
  {pages} {064042} (\bibinfo {year} {2017})}\BibitemShut {NoStop}%
\bibitem [{\citenamefont {Fourdrinoy}\ \emph {et~al.}(2022)\citenamefont
  {Fourdrinoy}, \citenamefont {Robertson}, \citenamefont {James}, \citenamefont
  {Fabbri},\ and\ \citenamefont {Rousseaux}}]{fourdrinoy2022correlations}%
  \BibitemOpen
  \bibfield  {author} {\bibinfo {author} {\bibfnamefont {Johan}\ \bibnamefont
  {Fourdrinoy}}, \bibinfo {author} {\bibfnamefont {Scott}\ \bibnamefont
  {Robertson}}, \bibinfo {author} {\bibfnamefont {Nicolas}\ \bibnamefont
  {James}}, \bibinfo {author} {\bibfnamefont {Alessandro}\ \bibnamefont
  {Fabbri}}, \ and\ \bibinfo {author} {\bibfnamefont {Germain}\ \bibnamefont
  {Rousseaux}},\ }\bibfield  {title} {\enquote {\bibinfo {title} {Correlations
  on weakly time-dependent transcritical white-hole flows},}\ }\href {\doibase
  10.1103/PhysRevD.105.085022} {\bibfield  {journal} {\bibinfo  {journal}
  {Phys. Rev. D}\ }\textbf {\bibinfo {volume} {105}},\ \bibinfo {pages}
  {085022} (\bibinfo {year} {2022})}\BibitemShut {NoStop}%
\bibitem [{\citenamefont {DeWitt}(1975)}]{dewitt1975quantum}%
  \BibitemOpen
  \bibfield  {author} {\bibinfo {author} {\bibfnamefont {Bryce~S}\ \bibnamefont
  {DeWitt}},\ }\bibfield  {title} {\enquote {\bibinfo {title} {Quantum field
  theory in curved spacetime},}\ }\href
  {https://www.sciencedirect.com/science/article/abs/pii/0370157375900514}
  {\bibfield  {journal} {\bibinfo  {journal} {Phys. Rep.}\ }\textbf {\bibinfo
  {volume} {19}},\ \bibinfo {pages} {295--357} (\bibinfo {year}
  {1975})}\BibitemShut {NoStop}%
\bibitem [{\citenamefont {Anderson}\ \emph {et~al.}(2013)\citenamefont
  {Anderson}, \citenamefont {Balbinot}, \citenamefont {Fabbri},\ and\
  \citenamefont {Parentani}}]{PhysRevD.87.124018}%
  \BibitemOpen
  \bibfield  {author} {\bibinfo {author} {\bibfnamefont {Paul~R.}\ \bibnamefont
  {Anderson}}, \bibinfo {author} {\bibfnamefont {Roberto}\ \bibnamefont
  {Balbinot}}, \bibinfo {author} {\bibfnamefont {Alessandro}\ \bibnamefont
  {Fabbri}}, \ and\ \bibinfo {author} {\bibfnamefont {Renaud}\ \bibnamefont
  {Parentani}},\ }\bibfield  {title} {\enquote {\bibinfo {title} {{Hawking}
  radiation correlations in {Bose}-{Einstein} condensates using quantum field
  theory in curved space},}\ }\href {\doibase 10.1103/PhysRevD.87.124018}
  {\bibfield  {journal} {\bibinfo  {journal} {Phys. Rev. D}\ }\textbf {\bibinfo
  {volume} {87}},\ \bibinfo {pages} {124018} (\bibinfo {year}
  {2013})}\BibitemShut {NoStop}%
\bibitem [{\citenamefont {Anderson}\ \emph {et~al.}(2014)\citenamefont
  {Anderson}, \citenamefont {Balbinot}, \citenamefont {Fabbri},\ and\
  \citenamefont {Parentani}}]{PhysRevD.90.104044}%
  \BibitemOpen
  \bibfield  {author} {\bibinfo {author} {\bibfnamefont {Paul~R.}\ \bibnamefont
  {Anderson}}, \bibinfo {author} {\bibfnamefont {Roberto}\ \bibnamefont
  {Balbinot}}, \bibinfo {author} {\bibfnamefont {Alessandro}\ \bibnamefont
  {Fabbri}}, \ and\ \bibinfo {author} {\bibfnamefont {Renaud}\ \bibnamefont
  {Parentani}},\ }\bibfield  {title} {\enquote {\bibinfo {title} {Gray-body
  factor and infrared divergences in 1{D BEC} acoustic black holes},}\ }\href
  {\doibase 10.1103/PhysRevD.90.104044} {\bibfield  {journal} {\bibinfo
  {journal} {Phys. Rev. D}\ }\textbf {\bibinfo {volume} {90}},\ \bibinfo
  {pages} {104044} (\bibinfo {year} {2014})}\BibitemShut {NoStop}%
\bibitem [{\citenamefont {Coutant}\ and\ \citenamefont
  {Parentani}(2014)}]{coutant2014hawking}%
  \BibitemOpen
  \bibfield  {author} {\bibinfo {author} {\bibfnamefont {Antonin}\ \bibnamefont
  {Coutant}}\ and\ \bibinfo {author} {\bibfnamefont {Renaud}\ \bibnamefont
  {Parentani}},\ }\bibfield  {title} {\enquote {\bibinfo {title} {Hawking
  radiation with dispersion: The broadened horizon paradigm},}\ }\href
  {\doibase 10.1103/PhysRevD.90.121501} {\bibfield  {journal} {\bibinfo
  {journal} {Phys. Rev. D}\ }\textbf {\bibinfo {volume} {90}},\ \bibinfo
  {pages} {121501} (\bibinfo {year} {2014})}\BibitemShut {NoStop}%
\bibitem [{\citenamefont {Bossard}\ \emph {et~al.}(2023)\citenamefont
  {Bossard}, \citenamefont {James}, \citenamefont {Aucouturier}, \citenamefont
  {Fourdrinoy}, \citenamefont {Robertson},\ and\ \citenamefont
  {Rousseaux}}]{BHLpreprint}%
  \BibitemOpen
  \bibfield  {author} {\bibinfo {author} {\bibfnamefont {Alexis}\ \bibnamefont
  {Bossard}}, \bibinfo {author} {\bibfnamefont {Nicolas}\ \bibnamefont
  {James}}, \bibinfo {author} {\bibfnamefont {Camille}\ \bibnamefont
  {Aucouturier}}, \bibinfo {author} {\bibfnamefont {Johan}\ \bibnamefont
  {Fourdrinoy}}, \bibinfo {author} {\bibfnamefont {Scott}\ \bibnamefont
  {Robertson}}, \ and\ \bibinfo {author} {\bibfnamefont {Germain}\ \bibnamefont
  {Rousseaux}},\ }\bibfield  {title} {\enquote {\bibinfo {title} {How to create
  analogue black hole or white fountain horizons and {LASER} cavities in
  experimental free surface hydrodynamics?}}\ }\href
  {https://arxiv.org/abs/2307.11022} {\bibfield  {journal} {\bibinfo  {journal}
  {arXiv:2307.11022}\ } (\bibinfo {year} {2023})},\ \bibinfo {note}
  {(preprint)}\BibitemShut {NoStop}%
\bibitem [{\citenamefont {Landau}\ and\ \citenamefont
  {Lifshitz}(2013)}]{landau_fluid_2013}%
  \BibitemOpen
  \bibfield  {author} {\bibinfo {author} {\bibfnamefont {Lev~Davidovich}\
  \bibnamefont {Landau}}\ and\ \bibinfo {author} {\bibfnamefont
  {Evgenii~Mikhailovich}\ \bibnamefont {Lifshitz}},\ }\href@noop {} {\emph
  {\bibinfo {title} {Fluid mechanics: Landau And Lifshitz: course of
  theoretical physics, Volume 6}}},\ Vol.~\bibinfo {volume} {6}\ (\bibinfo
  {publisher} {Elsevier},\ \bibinfo {year} {2013})\BibitemShut {NoStop}%
\bibitem [{\citenamefont {Weinfurtner}\ \emph {et~al.}(2013)\citenamefont
  {Weinfurtner}, \citenamefont {Tedford}, \citenamefont {Penrice},
  \citenamefont {Unruh},\ and\ \citenamefont
  {Lawrence}}]{weinfurtner2013classical}%
  \BibitemOpen
  \bibfield  {author} {\bibinfo {author} {\bibfnamefont {Silke}\ \bibnamefont
  {Weinfurtner}}, \bibinfo {author} {\bibfnamefont {Edmund~W.}\ \bibnamefont
  {Tedford}}, \bibinfo {author} {\bibfnamefont {Matthew C.~J.}\ \bibnamefont
  {Penrice}}, \bibinfo {author} {\bibfnamefont {William~G.}\ \bibnamefont
  {Unruh}}, \ and\ \bibinfo {author} {\bibfnamefont {Gregory~A.}\ \bibnamefont
  {Lawrence}},\ }\enquote {\bibinfo {title} {Classical aspects of {Hawking}
  radiation verified in analogue gravity experiment},}\ in\ \href {\doibase
  10.1007/978-3-319-00266-8_8} {\emph {\bibinfo {booktitle} {Analogue Gravity
  Phenomenology: Analogue Spacetimes and Horizons, from Theory to
  Experiment}}},\ \bibinfo {editor} {edited by\ \bibinfo {editor}
  {\bibfnamefont {Daniele}\ \bibnamefont {Faccio}}, \bibinfo {editor}
  {\bibfnamefont {Francesco}\ \bibnamefont {Belgiorno}}, \bibinfo {editor}
  {\bibfnamefont {Sergio}\ \bibnamefont {Cacciatori}}, \bibinfo {editor}
  {\bibfnamefont {Vittorio}\ \bibnamefont {Gorini}}, \bibinfo {editor}
  {\bibfnamefont {Stefano}\ \bibnamefont {Liberati}}, \ and\ \bibinfo {editor}
  {\bibfnamefont {Ugo}\ \bibnamefont {Moschella}}}\ (\bibinfo  {publisher}
  {Springer International Publishing},\ \bibinfo {address} {Cham},\ \bibinfo
  {year} {2013})\ pp.\ \bibinfo {pages} {167--180}\BibitemShut {NoStop}%
\bibitem [{\citenamefont {Euvé}(2017)}]{Euver2017}%
  \BibitemOpen
  \bibfield  {author} {\bibinfo {author} {\bibfnamefont {Léo-Paul}\
  \bibnamefont {Euvé}},\ }\emph {\bibinfo {title} {Interactions
  ondes-courant-obstacle : application à la physique des trous noirs}},\ \href
  {http://www.theses.fr/2017POIT2280} {Ph.D. thesis} (\bibinfo {year} {2017}),\
  \bibinfo {note} {thèse de doctorat dirigée par Huberson, Serge et
  Rousseaux, Germain Mécanique des fluides Poitiers 2017}\BibitemShut
  {NoStop}%
\bibitem [{\citenamefont {Peskin}(2018)}]{Peskin:1995ev}%
  \BibitemOpen
  \bibfield  {author} {\bibinfo {author} {\bibfnamefont {Michael~E}\
  \bibnamefont {Peskin}},\ }\href@noop {} {\emph {\bibinfo {title} {An
  introduction to quantum field theory}}}\ (\bibinfo  {publisher} {CRC press},\
  \bibinfo {year} {2018})\BibitemShut {NoStop}%
\bibitem [{\citenamefont {Baird}(1970)}]{baird1970new}%
  \BibitemOpen
  \bibfield  {author} {\bibinfo {author} {\bibfnamefont {L.~C.}\ \bibnamefont
  {Baird}},\ }\bibfield  {title} {\enquote {\bibinfo {title} {New integral
  formulation of the {Schr{\"o}dinger} equation},}\ }\href
  {https://pubs.aip.org/aip/jmp/article-abstract/11/8/2235/388150/New-Integral-Formulation-of-the-Schrodinger}
  {\bibfield  {journal} {\bibinfo  {journal} {J. Math. Phys.}\ }\textbf
  {\bibinfo {volume} {11}},\ \bibinfo {pages} {2235--2242} (\bibinfo {year}
  {1970})}\BibitemShut {NoStop}%
\bibitem [{\citenamefont {Massar}\ and\ \citenamefont
  {Parentani}(1998)}]{massar1998particle}%
  \BibitemOpen
  \bibfield  {author} {\bibinfo {author} {\bibfnamefont {S.}~\bibnamefont
  {Massar}}\ and\ \bibinfo {author} {\bibfnamefont {R.}~\bibnamefont
  {Parentani}},\ }\bibfield  {title} {\enquote {\bibinfo {title} {Particle
  creation and non-adiabatic transitions in quantum cosmology},}\ }\href
  {https://www.sciencedirect.com/science/article/abs/pii/S0550321397007189}
  {\bibfield  {journal} {\bibinfo  {journal} {Nucl. Phys. B}\ }\textbf
  {\bibinfo {volume} {513}},\ \bibinfo {pages} {375--401} (\bibinfo {year}
  {1998})}\BibitemShut {NoStop}%
\end{thebibliography}%

\newpage

\section*{Supplemental Material}

\section{Equations of motion}
\label{app:em}

In this section, we shall provide some details of the derivation of the equations of motion, Eqs.~\eqref{eqs:mot}.  Throughout, we shall use the superscripts $+$ and $-$ to indicate quantities which are linked to the upper layer and the lower layer, respectively.

\subsection{Basic equations}

Since the flow is incompressible, the fluid density $\rho$ remains constant, and the continuity equation becomes a  constraint on the flow velocity $\textbf{V}$: 
\begin{equation}
    \boldsymbol{\nabla}\cdot\textbf{V}=\partial_x u+\partial_z w=0.
    \label{eq:incompressibility}
\end{equation}
As we assume an inviscid flow, it is governed by the Euler equations~\cite{landau_fluid_2013}
\begin{equation}
    \frac{d\textbf{V}}{dt}=\partial_t\textbf{V}+(\textbf{V}\cdot\boldsymbol{\nabla})\textbf{V}=-\frac1\rho\boldsymbol{\nabla} P-\textbf{g},
    \label{eq:euler}
\end{equation}
where $P$ is the pressure and $\textbf{g}=g\hat{\textbf{z}}$ represents the gravitational acceleration.

As discussed in the main text, we consider a 2D flow consisting of two layers: an upper layer with zero vorticity, and a lower layer of constant vorticity.  Since, in a 2D flow, vorticity is propagated along streamlines ({\it i.e.}, we have $\left(\partial_{t}+\textbf{V} \cdot \boldsymbol{\nabla} \right)\boldsymbol{\Omega} = \textbf{0}$), then in the presence of a wave we will continue to have a division into two layers, though the boundary may fluctuate.
Therefore, we can write
\begin{equation}
    \boldsymbol{\nabla}\times\textbf{V}=
    \begin{cases}
    \textbf{0} & b+H_\Omega\le z \le b+H\\
    \hat{\textbf{y}}\Omega & b\le z\le b+H_\Omega
    \end{cases},
\end{equation}
where $H_{\Omega}$ and $H$ are the space- and time-dependent heights of the corresponding boundaries (see Fig.~\ref{fig:system}).
Note in particular that, since the vorticity vanishes in the upper layer, we can introduce a velocity potential $\varphi^+$ such that $u^+=-\partial_x\varphi^+$.

Boundary conditions are imposed on the bottom ($z=b$), on the free surface ($z=b+H$), and on the interface between the two layers ($z=b+H_{\Omega}$). 
At each boundary, there is no time-dependence if the velocity is parallel to the boundary, {\it i.e.}, if the vertical component of the flow velocity is exactly what is needed to keep the flow tangent to the boundary.  Any additional contribution to the vertical component thus induces a time-variation of the position of the boundary.  Since the bottom is necessarily fixed, there can be no such additional term, and the flow velocity must be exactly tangent to the bottom.  In addition to these kinematical constraints, there is an additional boundary condition in the form of the constancy of the pressure at the free surface; it can be set equal to the atmospheric pressure, but since the equations are invariant under constant shifts of the pressure, it is most straightforward to simply set it to zero.
In mathematical terms, these boundary conditions are
\begin{align}
    & w|_{z=b}=u|_{z=b} \, \partial_xb,\qquad\quad P|_{z=b+H}=0, \nonumber \\[0.3cm]
    & w|_{z=b+H}=u|_{z=b+H} \, \partial_x(b+H)+\partial_tH, \nonumber \\[0.3cm]
    & w|_{z=b+H_{\Omega}}=u|_{z=b+H_{\Omega}} \, \partial_x(b+H_\Omega)+\partial_tH_\Omega.
\end{align}

\subsection{Slowly-varying limit}

We now take the limit in which variations in the longitudinal ($\hat{x}$) direction (as well as those in time) take place on much longer scales than variations in the vertical ($\hat{z}$) direction.  To implement this, we introduce a dimensionless scaling parameter $\epsilon \ll 1$ that acts as if the system were stretched along $x$, and we make the following replacements:
\begin{equation}
    \label{eq:lwl}
    \partial_t,\ \partial_x,\ w\ \to\ \varepsilon\partial_t,\ \varepsilon\partial_x,\ \varepsilon w \,.
\end{equation}
These are then substituted into the above equations, and we keep only the lowest-order terms in $\epsilon$.  This simplifies two of the equations.  First, the $z$-component of the Euler equations becomes
\begin{equation}
    \frac{1}{\rho} \partial_{z}P + g = 0 \,,
\end{equation}
which combined with the vanishing of the pressure at the free surface immediately implies that $P = \rho g \left(b+H-z\right)$.  Second, the vorticity, which in 2D has only a single component (directed along $y$), takes the form
\begin{equation}
    \left(\boldsymbol{\nabla} \times \textbf{V}\right)_{y} = \partial_{z}u \,,
\end{equation}
with no remaining contribution from the vertical component $w$.  Given the two-layer form of the vorticity that we are considering, this immediately implies that $u^{+}$ is $z$-independent while $u^{-} = u^{+} + \Omega \left(z-b-H_{\Omega}\right)$ (by continuity of $u$ at the interface between the upper and lower layers).  In turn, this implies that the $x$-derivative of $u$ is $z$-independent in both layers, and hence (through the incompressibility condition~(\ref{eq:incompressibility})) that $w$ varies linearly with $z$ in each layer.

\subsection{Separation into background and perturbations}

The Analogue Gravity viewpoint relies on our ability to separate the system into two components of very different strengths.  One is the strong background, which plays the role of the effective spacetime, and the other is the perturbations on top of this background which are meant to be so small that any backreaction on the background can be neglected.

Here, we separate the system into a time-independent background flow (quantities associated with the background will typically be indicated with the subscript $0$), plus a time-dependent perturbation of very small amplitude (these will typically be indicated by the use of the prefix $\delta$).  All of the equations listed above are linearized in the amplitude of the perturbations, yielding two distinct sets of equations: a set of nonlinear equations for the background, and a set of linear equations for the perturbations.

\subsection{Background}

The equations for the background are essentially found by setting all time-derivatives to zero.  They can be summarized by the following equations, indicating that the background is characterised by three conserved quantities:
\begin{eqnarray}
    \partial_{x} \left[ u_{0}^{+} \left( h - h_{\Omega} \right) \right] &=& 0 \,, \nonumber \\[0.1cm]
    \partial_{x} \left[ \left( u_{0}^{+} - \frac{1}{2} \Omega h_{\Omega} \right) h_{\Omega} \right] &=& 0 \,, \nonumber \\[0.1cm]
    \partial_{x} \left[ \frac{1}{2g} \left(u_{0}^{+}\right)^{2} + h + b \right] &=& 0 \,.
\end{eqnarray}
The first two equations are derived by imposing the kinematical boundary conditions at each boundary with $w$ varying linearly in $z$ within each layer.  They express the fact that the mass flux rate is conserved separately within each layer.  The third equation is found from the $x$-component of the Euler equations with the pressure $P = \rho g \left(h + b - z\right)$; it is just Bernoulli's equation applied to the streamline at the free surface.

While any background flow is characterised by these three conserved quantities, if we impose that the flow be {\it transcritical} then we require only two conserved quantities, the other being fixed by the transcriticality condition.  For instance, we may set $q^{+} = u_{0}^{+}\left(h-h_{\Omega}\right)$ and $q^{-} = \left(u_{0}^{+}-\Omega h_{\Omega}/2\right) h_{\Omega}$, and this allows us to write $h$ in terms of $h_{\Omega}$:
\begin{equation}
    h = h\left(h_{\Omega}\right) = h_{\Omega} \left(1 + \frac{q^{+}}{q_{-} + \frac{1}{2} \Omega h_{\Omega}^{2}} \right) \,.
\end{equation}
Using this, the Bernoulli equation can itself be expressed in terms of $h_{\Omega}$ alone:
\begin{equation}
    e + b = {\rm const.} \,, \qquad e = \frac{\left(q^{+}\right)^{2}}{2g \left(h\left(h_{\Omega}\right)-h_{\Omega}\right)^{2}} + h\left(h_{\Omega}\right) \,.
\end{equation}
The function $e\left(h_{\Omega}\right)$ diverges as $h_{\Omega} \to 0$ and as $h_{\Omega} \to \infty$, and exhibits a minimum at some finite $h_{\Omega}$.  It can be shown that the flow is subcritical for $h_{\Omega}$ larger than this critical value, and supercritical for $h_{\Omega}$ smaller than this critical value; indeed, the Bernoulli equation implies that the energy of the flow is mostly potential for large $h_{\Omega}$, and mostly kinetic for small $h_{\Omega}$.  The constancy of $e+b$ determines how $h_{\Omega}$ (and, in turn, all other quantities) varies with position, once a particular bottom profile $b(x)$ is given.  However, for most values of $e+b$, this will typically yield two disconnected solutions, one on either side of the minimum of $e\left(h_{\Omega}\right)$.  In order for transcriticality to occur, the solution must pass from one solution branch to the other, and the only way for this to occur smoothly is for it to pass through the minimum of $e\left(h_{\Omega}\right)$ at precisely the top of the obstacle, where $b=b_{\rm max}$.  Therefore, for a transcritical flow, the constant must be set equal to $e_{\rm min} + b_{\rm max}$.

The method just described is used to determine the background in Fig.~\ref{fig:systemsc} in the main text. In particular, for this system, we choose (inspired by experimental data) 
the fluxes $q_+=0.08 \, {\rm m}^{2}/{\rm s}$ and $q_-=0.07 \, {\rm m}^{2}/{\rm s}$, and a vorticity $\Omega=4.6 \, {\rm s}^{-1}$. For the bottom we have considered the obstacle used in the Vancouver experiment~\cite{weinfurtner2013classical, Euver2017}, which has the following form:
\begin{equation}
    \frac{b(x)}{1 \, {\rm m}} =
    \begin{cases}
    f(x,x_1) & x_1\le x \le x_2\\[0.1cm]
    0.1 & x_2<x\le x_3\\[0.1cm]
    f(x,x_1-x_2) & x_3< x \le x_4\\[0.1cm]
    g(x,x_1-x_2) & x_4< x \le x_5\\[0.1cm]
    \end{cases},
\end{equation}
where $x_1=-0.45\ \mathrm{m}$, $x_2=-0.15\ \mathrm{m}$, $x_3=0\ \mathrm{m}$, $x_4=0.34\ \mathrm{m}$ and $x_5=1.15\ \mathrm{m}$, while functions $f(x,y)$ and $g(x,y)$ are defined as
\begin{align}
    &f(x,y)=2a(1-(x-y)-e^{-\kappa(x-y)}),\\[0.1cm]
    &g(x,y)=f(x_4,y)-(x-x_4)\tan(\alpha),
\end{align}
with the parameters $a=0.094\ \mathrm{m}$, $\kappa=5.94\ \mathrm{m}^{-1}$ and $\alpha = 4.5^{\circ}$.

\subsection{Perturbations}

From the $x$-component of Eq.~\eqref{eq:euler}, we find
\begin{eqnarray}
    \left[\partial_{t} + \partial_{x}u_{0}^{+}\right] \delta u^{+} &=& -\frac1\rho \partial_{x}\left(\delta P^{+}\right) \,, \nonumber \\
    \left[\partial_{t} + \partial_{x}\left(u_{0}^{+}-\Omega h_{\Omega}\right)\right] \delta u^{-} &=& -\frac1\rho \partial_{x}\left(\delta P^{-}\right) \,.
    \label{eq:pert1}
\end{eqnarray}
The expression for the pressure, $P = \rho g \left( b + H - z \right)$, holds in both layers and can be immediately linearized to give
\begin{equation}
    \delta P = \rho g \eta \,,
    \label{eq:pressure}
\end{equation}
so that the right-hand side of Eqs.~(\ref{eq:pert1}) becomes simply $-g\partial_{x}\eta$.  Replacing $\delta u^{\pm} = -\partial_{x}\left(\delta \varphi^{\pm}\right)$, we can integrate Eqs.~(\ref{eq:pert1}) with respect to $x$, yielding:
\begin{eqnarray}
    \left[\partial_{t} + u_{0}^{+}\partial_{x}\right] \delta\varphi^{+} &=& g\eta \,, \nonumber \\
    \left[\partial_{t} + \left(u_{0}^{+}-\Omega h_{\Omega}\right)\partial_{x}\right]\delta\varphi^{-} &=& g\eta \,.
\end{eqnarray}
Imposing the boundary conditions gives
\begin{eqnarray}
    \left[\partial_{t} + \partial_{x}u_{0}^{+}\right]\eta_{\Omega} &=& - \partial_{x}\left[h_{\Omega}\,\delta u^{-}\right] \,, \nonumber \\
    \left[\partial_{t} + \partial_{x}u_{0}^{+}\right]\left(\eta-\eta_{\Omega}\right) &=& -\partial_{x}\left[\left(h-h_{\Omega}\right) \delta u^{+}\right] \,.
\end{eqnarray}
These four equations can be combined to give the equations of motion we are seeking.  Note that, in the main text, we introduce $\delta\varphi^{\Omega} = \delta\varphi^{+}-\delta\varphi^{-}$, so that $\delta u^{+}-\delta u^{-} = \Omega \eta_{\Omega} = -\partial_{x}\delta\varphi^{\Omega}$.
At the end, we are left with Eqs.~(\ref{eqs:mot}) and~(\ref{eqs:free_surface_deformations}) of the main text.

\section{Elementary excitations}
\label{app:modes}

\subsection{Dispersion relation}
\label{subsec:disp_rel}

The dispersion relation is found by taking a constant background and assuming a plane wave solution such that both $\delta\varphi^{+} = A^{+} e^{ikx-i\omega t}$ and $\delta\varphi^{\Omega} = A^{\Omega} e^{ikx-i\omega t}$, then solving for the necessary relationship between $\omega$ and $k$.  The wave equations~(\ref{eqs:mot}) yield the linear system
\begin{equation}
\left[\begin{array}{cc} \left(\omega-u_{0}^{+}k\right)^{2} - gh \, k^{2} & -g h_{\Omega} \, k^{2} \\ \Omega h_{\Omega} \, k^{2} & \left(\omega-\left(u_{0}^{+}-\Omega h_{\Omega}\right)k\right) k \end{array} \right] \left( \begin{array}{c} A^{+} \\ A^{\Omega} \end{array} \right) = \left( \begin{array}{c} 0 \\ 0 \end{array} \right) \,.
\label{eq:matrix}
\end{equation}
In order for this system to have a non-trivial solution, the determinant of the matrix on the left-hand side must vanish.  It is this that yields the dispersion relation, which is given in Eq.~(\ref{eq:disp_rel}) in the main text in the form of a cubic equation for the phase velocity $v_{p} = \omega/k$.  Therefore, there are generally three different branches of the dispersion relation.

Taking the zero-vorticity limit $\Omega = 0$ of Eq.~(\ref{eq:disp_rel}), we have
\begin{equation}
    \left(v_{p}-u_{0}\right)^{3} - g h \left(v_{p}-u_{0}\right) = 0 \,,
\end{equation}
where we have removed the $+$ superscript from $u_{0}$ because there is only a single layer in this limit.
Here, $v_{p}-u_{0}$ is just the phase velocity $c$ of the wave with respect to the fluid, and we find that there are two solutions satisfying $c^{2} = g h$; these are precisely the two surface modes we are used to.  The third solution (the ``vorticity mode'') is trivial here, with $c=0$; it has no relevance in this limit, but becomes a propagating mode as soon as the vorticity layer is introduced.

We can deduce certain properties of the solutions of the dispersion relation by studying the coefficients of the cubic equation, since we know that, by factorization,
\begin{equation}
    x^{3} + a x^{2} + b x + c  =  \left(x-x_{1}\right) \left(x-x_{2}\right) \left(x-x_{3}\right)
\end{equation}
where the $x_{j}$ are the roots of the polynomial, and therefore
\begin{eqnarray}
a &=& = - \left(x_{1}+x_{2}+x_{3}\right) \,, \nonumber \\
b &=& x_{1}x_{2} + x_{2}x_{3} + x_{3}x_{1} \,, \nonumber \\
c &=& - x_{1}x_{2}x_{3} \,.
\end{eqnarray}
For instance, Eq.~(\ref{eq:disp_rel}) is written such that the quadratic term in the cubic equation vanishes, meaning that the sum of the three solutions must also vanish.  We can re-write the cubic so that it is adapted to the flow in the upper layer:
\begin{equation}
    \begin{split}
        &(v_p-u_0^+)^3+\Omega h_\Omega(v_p-u_0^+)^2+\\[0.1cm]
        &-gh(v_p-u_0^+)-g\Omega h_\Omega(h-h_\Omega)=0.
    \end{split}
    \label{eq:toplayera}
\end{equation}
Focusing on the constant term, we may conclude that the product of the the three roots $v_{p}-u_{0}^{+}$ must be positive.  Since the two surface waves have opposite phase velocities with respect to the flow, this means that the vorticity mode must have a negative value of $v_{p}-u_{0}^{+}$, {\it i.e.}, the vorticity wave always propagates at a speed slower than the flow at the free surface.
Similarly, we can rewrite the cubic equation so that it is adapted to the flow speed on the bottom of the channel, $u_{0}^{b} = u_{0}^{+}-\Omega h_{\Omega}$: 
\begin{equation}
    \begin{split}
        &(v_p-u_0^b)^3-2(v_p-u_0^b)^2(u_0^+-u_0^b)+\\[0.1cm]
        &+(v_p-u_0^b)((u_0^+-u_0^b)^2-gh)+g\Omega h_\Omega^2=0.
    \end{split}
\end{equation}
In this case, the unknown in $v_p-u_0^b$, thus the phase velocity in the frame for which the speed of the bottom is null. With a reasoning analogous to the one used before, we find that the vorticity mode has a phase velocity which is faster than the speed of current on the bottom. 
In conclusion, the speed of the vorticity mode lies somewhere in between the flow speed on the bottom of the channel and the flow speed on the free surface.

\subsection{Effective phase velocity and flow speed}
\label{subsec:eff_velocities}

\begin{figure}
    \includegraphics[width=0.8\linewidth]{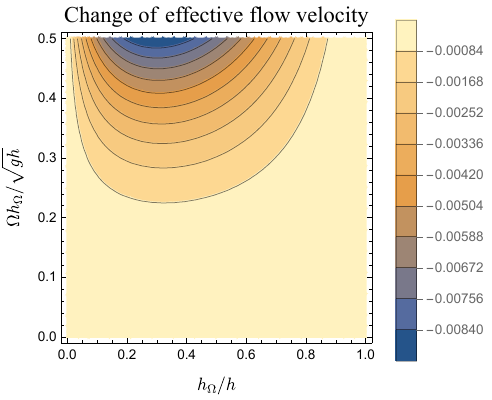}
    \includegraphics[width=0.8\linewidth]{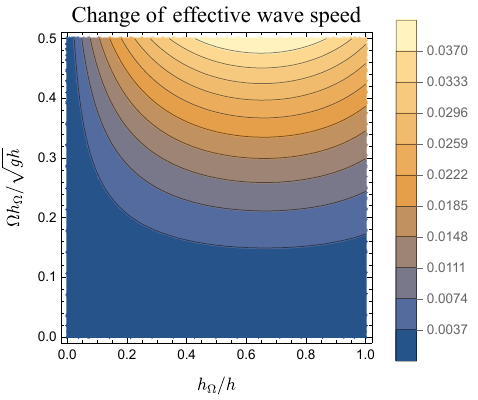}
    \caption{Top: Contour plot of the change in the effective flow velocity, defined as the difference between the sum of the co-current mode and the counter-current mode, and the average speed of the current in the background. Bottom: Contour plot of the change in the effective wave speed, defined as the difference between the co-current mode and the counter-current mode, and the wave speed in the absence of vorticity. Both plots are shown as functions of $h_\Omega/h$ and $\Omega h_\Omega/\sqrt{gh}$, where $\Omega h_\Omega=\Delta u$ is the difference between the speed at the top and the bottom.}
    \label{fig:effective}
\end{figure}

In an Analogue Gravity context, it is typically the propagation of the two surface modes that is of interest, these being of the form $v_{p} = u \pm c$ where $u$ is the flow velocity and $c$ is the wave speed with respect to the fluid.  If we neglect the vorticity mode, we may use the velocities of the two surface modes to derive an effective background by identifying them with $u_{\rm eff} \pm c_{\rm eff}$; then we have, by definition,
\begin{equation}
    u_{\rm eff} = \frac{1}{2}\left(v_{p}^{\rightarrow}+v_{p}^{\leftarrow}\right) \,, \qquad c_{\rm eff} = \frac{1}{2}\left(v_{p}^{\rightarrow}-v_{p}^{\leftarrow}\right) \,,
\end{equation}
where the arrow shows the direction of the corresponding surface wave.

The results are shown in Fig.~\ref{fig:effective}.
As independent variables, we use the frame-independent quantities $h_{\Omega}/h$ (the relative size of the vorticity layer) and $\Omega h_{\Omega} / \sqrt{g h}$ (the adimensionalised velocity difference across the vorticity layer).
In the first panel, we show the difference between the effective flow $u_{\rm eff}$ and the depth-averaged flow velocity
\begin{equation}
    \bar{u} = \frac{1}{h} \int_{0}^{h} dz \, u(z) = u_{0}^{+} - \frac{1}{2} \Omega \frac{h_{\Omega}^{2}}{h} \,,
\end{equation}
while in the second panel we show the difference between the effective wave speed $c_{\rm eff}$ and the expected wave speed $\sqrt{g h}$.
Both differences are adimensionalised by the frame-independent velocity $\sqrt{g h}$.

The most important observation is the smallness of the differences in the region of parameter space considered.  The change in the effective wave speed is far more significant than the change in the effective flow velocity, and tends to increase a little with respect to the expected value.

\subsection{Comparison with smooth velocity profile}

\begin{figure}
    \includegraphics[width=0.9\linewidth]{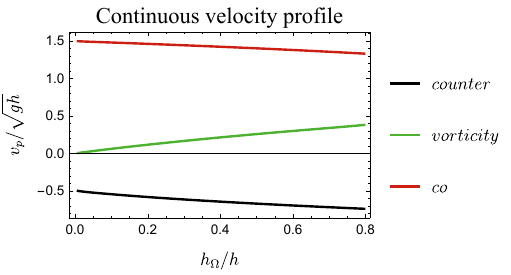}
    \includegraphics[width=0.9\linewidth]{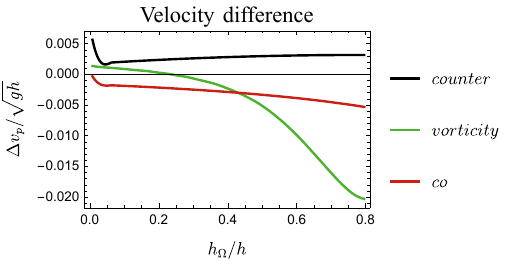}
    \caption{Top: Phase velocities for a model with continuous vorticity and a flat bottom, \emph{i.e.}, with an \( x \)-independent background, calculated numerically for a small wavenumber \( k \). The plot corresponds to the velocity profile given by Eq.~\eqref{eq:velprofile}, with \( u_0^{\text{top}}/\sqrt{gh} = 1/2 \), \( n = 6 \), and \( h_\Omega/h \) values ranging from 0.008 to 0.8. Bottom: Difference between the phase velocities for the model with continuous vorticity and the model with discontinuous vorticity as a function of $h_\Omega/h$. For this comparison, the same parameters were used, and the mode most affected is the vorticity mode (green). This phenomenon is attributed to the fact that the vorticity mode ``lives'' on the interface between the two layers, making it sensitive to the sharpness of the velocity profile.}
    \label{fig:differencevel}
\end{figure}

To dispel the possibility that the vorticity mode might be a mathematical artefact due to the idealisation of a discontinuous vorticity, we briefly consider the solutions of the dispersion relation on a background with a smooth velocity profile.
We consider the following:
\begin{equation}
    u_0(z)=u_0^{\text{top}}\left[\tanh\left(\frac{z}{L}\right)^n\right]^{1/n},
    \label{eq:velprofile}
\end{equation}
where $u_0^{\text{top}}$ is the asymptotic value, $L$ is a characteristic length scale on the order of $h_{\Omega}$, 
and $n$ is a parameter that controls the sharpness of the transition.  Exploiting the equations shown in~\cite{thompson1949propagation} and taking the limit $k \to 0$ (since~\cite{thompson1949propagation} includes dispersive effects that we have ignored here), 
we can obtain the values admitted for the phase velocity.
Typical results are shown in Fig.~\ref{fig:differencevel}, 
where we have chosen $u_0^{\text{top}}/\sqrt{gh} = 1/2$, $n=6$, and the values of $h_\Omega/h$ were between $0.008$ and $0.8$.
Importantly, there are indeed three non-trivial solutions, indicating that the presence of the vorticity mode is not an artefact of the discontinuity in the vorticity.

We may also ask to what extent the smoothing of the velocity profile affects the solutions.
To answer this, we compare the two models in the following way. We fix the values for $u_0^{\text{top}}$ and $n$, and vary the parameter $L$. For each value of $L$, we have a different velocity profile, and from it, we construct a linearised model that is comparable with the model with continuous vorticity: $\Omega$ is set equal to $u'(z=0)$, $u_{0}^{+}$ is set equal to $u_{0}^{\rm top}$, and $h_{\Omega}$ is defined as the point at which the linear and constant pieces of the profile match. 
Once this is done, it is easy to calculate the phase velocity for both models. 
In the lower panel of Figs.~\ref{fig:differencevel}, we show the difference in the phase velocities for the continuous and discontinuous models of vorticity, as a function of $h_\Omega/h$. For this case, we have chosen $u_0^{\text{top}}/\sqrt{gh} = 1/2$, $n=6$, and the values of $h_\Omega/h$ were between $0.008$ and $0.8$. We notice that the mode most affected is the vorticity mode (green). This phenomenon could be due to the fact that the vorticity mode ``lives'' on the interface between the two layers, making it the most sensitive to the sharpness (or lack thereof) of the velocity profile.

\section{Lagrangian description}
\label{app:canfield}

In the system we are considering, we have defined two potentials which are analogous to two massless scalar fields. We want to define the canonical fields and their conjugated momenta. To do this, we compare the physical momentum and energy of the system to those associated with the scalar fields according to standard field theory. 

\subsection{Momentum and energy}

If we assume the existence of two canonical scalar fields, $\phi^+$ and $\phi^\Omega$, with associated canonical momenta $\pi^+$ and $\pi^\Omega$, the physical momentum and the energy associated to these fields are~\cite{Peskin:1995ev}
\begin{align}
    \label{eq:momentumfield}
    &P^x=-\pi^+\partial_x\phi^+-\pi^\Omega\partial_x\phi^\Omega,\\[0.1cm]
    \label{eq:energyfield}
    &E=\pi^+\partial_t\phi^++\pi^\Omega\partial_t\phi^\Omega-\mathcal{L},
\end{align}
with $\mathcal{L}$ the total Lagrangian.

Considering the channel flow and the potentials $\delta\varphi^+,\ \delta\varphi^\Omega$, the total momentum and energy densities (in the $(x,y)$-plane) are defined as
\begin{eqnarray}
    P &=& \int_0^H\rho u\ dz \,, \nonumber \\
    E &=& \int_0^H\rho\left[\frac12 (u^2+w^2)+gz\right]\ dz.
\end{eqnarray}
To extract the corresponding densities associated with a particular plane wave, we subtract the densities that are present in the absence of any wave (and are thus associated with the background):
\begin{eqnarray}
    P_{\rm wave} &=& \rho \left[ \int_0^{h+\eta} \left(u_{0}+\delta u\right)\ dz - \int_0^{h} u_{0}\ dz \right] \,, \nonumber \\
    E_{\rm wave} &=& \rho\left[ \int_0^{h+\eta}\left(\frac12 \left(u_{0}+\delta u\right)^2+gz\right)\ dz \right. \nonumber \\
    && \qquad \qquad \qquad \left. - \int_0^{h}\left(\frac12 u_{0}^2+gz\right)\ dz \right] \,,
\end{eqnarray}
where we have neglected the contribution of $w^{2}$ to the energy density because we work here in the long-wavelength limit.
We thus find
\begin{align}
    \label{eq:momentumwave}
    P_{\rm wave} =& \rho\left[\eta \, \delta u^{+} -\frac{\Omega}{2}\eta_{\Omega}^{2} \right] \,, \\
    \label{eq:energywave}
    E_{\rm wave} =& \rho \left[\frac{1}{2} h(\delta u^{+})^{2}+\frac{1}{2} g \eta^{2} + \frac{1}{2} \Omega^{2} h_{\Omega} \eta_{\Omega}^{2} \right. \nonumber \\
    & \qquad\qquad \left. + \Omega h_{\Omega} \, \eta_{\Omega} \, \delta u^{+} \right] + u_{0}^{+} P_{\rm wave} \,,
\end{align}
where we have neglected the contribution of linear terms because their average is zero.

\subsection{Canonical fields and Lagrangian}

Comparing Eqs.~\eqref{eq:momentumfield} and~\eqref{eq:energyfield} with Eqs.~\eqref{eq:momentumwave} and~\eqref{eq:energywave}, and using the relations~(\ref{eqs:free_surface_deformations}) between $\eta$, $\eta_{\Omega}$ and the scalar fields $\delta\varphi^{+}$, $\delta\varphi^{\Omega}$, we are led to define the canonical fields and their conjugated momenta as follows:
\begin{align}
    \label{eq:apiu}
    &\phi^+=\sqrt{\frac{\rho}{g}}\delta\varphi^+,\quad \pi^+=\sqrt{\rho g}\eta=(\partial_t+u_0^+\partial_x)\phi^+,\\[0.1cm]
    \label{eq:omega}
    &\phi^\Omega=\sqrt{\frac{\rho}{g}}\delta\varphi^\Omega,\quad \pi^\Omega=-\frac{\sqrt{\rho g}}{2}\eta_\Omega=\frac{g}{2\Omega}\partial_x\phi^\Omega.
\end{align}
From these definitions, and from standard results of field theory, we get an effective Lagrangian which takes the form
\begin{equation}
    \mathcal{L} = \mathcal{L}_{+} + \mathcal{L}_{\Omega} + \mathcal{L}_{\rm int} \,,
\end{equation}
where $\mathcal{L}_{+}$ and $\mathcal{L}_{\Omega}$ are the Lagrangians of the ``free'' fields $\delta\varphi_{+}$ and $\delta\varphi_{\Omega}$, and $\mathcal{L}_{\rm int}$ describes the effective interaction between the two fields.  We find
\begin{align}
    \mathcal{L}_{+} &= \frac{1}{2 }\Big[((\partial_t+\unotp\partial_x)\phi^+)^2-gh(\partial_x\phi^+)^2\Big] \,, \nonumber \\
    \mathcal{L}_{\Omega} &= \frac{g}{2\Omega}\Big[ \partial_t\phi^\Omega\partial_x\phi^\Omega+(\unotp-\Omega h_\Omega)(\partial_x\phi^\Omega)^2 \Big] \,, \nonumber \\
    \mathcal{L}_{\rm int} &= g \, h_\Omega\partial_x\phi^\Omega\partial_x\phi^+ \,,
    \label{eqs:Lagrangians}
\end{align}
where $\rho$ is the density of the fluid (this factor having been included to give $\mathcal{L}$ the appropriate dimensions of energy per unit area).  The factor of $1/\Omega$ in $\mathcal{L}_{\Omega}$ might seem strange as it diverges in the limit $\Omega \to 0$, but this succeeds in making $\phi^{\Omega}$ more difficult to excite and thus to ensure continuity of $\delta u$ in the absence of vorticity.  It is also to be noted that, with the Lagrangian given by Eqs.~(\ref{eqs:Lagrangians}), the free-surface deformations of Eqs.~(\ref{eqs:free_surface_deformations}) are simply proportional to the corresponding canonical momenta.

\subsection{Scalar product and WKB modes}
\label{app:wkbsols}

With the canonical fields at our disposal, we may immediately calculate the conserved scalar product between any two (complex) solutions of the wave equations.
Given the expression for the canonical fields and the conjugated momenta (Eqs.~\eqref{eq:apiu},~\eqref{eq:omega}), we define the vectors
\begin{equation}
    \boldsymbol{\phi}=
    \begin{pmatrix}
        \phi^+\\[0.1cm]
        \phi^\Omega
    \end{pmatrix},
    \qquad\quad
    \boldsymbol{\pi}=
    \begin{pmatrix}
        \pi^+\\[0.1cm]
        \pi^\Omega
    \end{pmatrix}.
    \label{eq:vectors}
\end{equation}
Denoting the scalar product using the standard bracket notation $(\ ,\ )$, then given two solutions $\boldsymbol{\phi}_{1}$ and $\boldsymbol{\phi}_{2}$ with corresponding momenta $\boldsymbol{\pi}_{1}$ and $\boldsymbol{\pi}_{2}$, we have
\begin{equation}
    (\boldsymbol{\phi}_1,\boldsymbol{\phi}_2)=i\int_{-\infty}^{+\infty}dx(\boldsymbol{\phi}_1^*\cdot\boldsymbol{\pi}_2-\boldsymbol{\phi}_2\cdot\boldsymbol{\pi}_1^*).
\end{equation}
The {\it norm} of a particular solution is just $\left(\boldsymbol{\phi},\boldsymbol{\phi}\right)$.
With this we can define a set of normalized modes with unit incident norm.
Moreover, the normalization allows us to define the WKB modes of the system~\cite{baird1970new,massar1998particle}, which become exact solutions in the limit of a slowly-varying background and are helpful when we come to define scattering coefficients. 
In particular, we can write 
\begin{equation}
    \boldsymbol{\phi}_\omega=
    \begin{pmatrix}
        \phi_\omega^+\\[0.1cm]
        \phi_\omega^\Omega
    \end{pmatrix}
    = \mathcal{N} \, e^{-i\omega t+i\int^x k(x')dx'}
    \begin{pmatrix}
        \sin\theta\\[0.1cm]
        \cos\theta
    \end{pmatrix}.
    \label{eq:wkb}
\end{equation}
The vector $\left({\rm sin}\theta \,, {\rm cos}\theta\right)$ is just a solution of Eq.~(\ref{eq:matrix}), parameterised by the angle $\theta$ which satisfies
\begin{equation}
    \tan\theta=\frac{gh_\Omega}{(v_p-\unotp)^2-gh}=\frac{\unotp-v_p-\Omega h_\Omega}{\Omega h_\Omega}.
\end{equation}
For $\boldsymbol{\phi}_{\omega}$ to be normalized requires
\begin{equation}
    \mathcal{N} =\left|\left[(v_{p}-\unotp)\sin^2\theta-\frac{g}{2\Omega}\cos^2\theta\right] \omega \right|^{-1/2}.
\end{equation}
Note that, in the case without vorticity ($\Omega=0$), only $\phi^{+}$ is excited so we have $\sin \theta = 1$ and $\cos \theta = 0$.  Then $\mathcal{N} = 1/\sqrt{c \, \omega}$ where $c$ is the surface wave speed with respect to the fluid.

\section{Scattering amplitudes}
\label{app:sa}

\begin{figure}

    \includegraphics[width=0.9\linewidth]{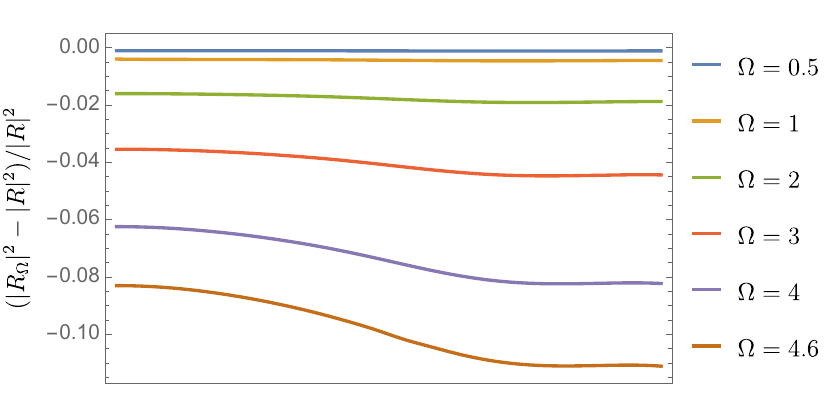}
    
    \vspace{0.2cm}
    
    \includegraphics[width=0.9\linewidth]{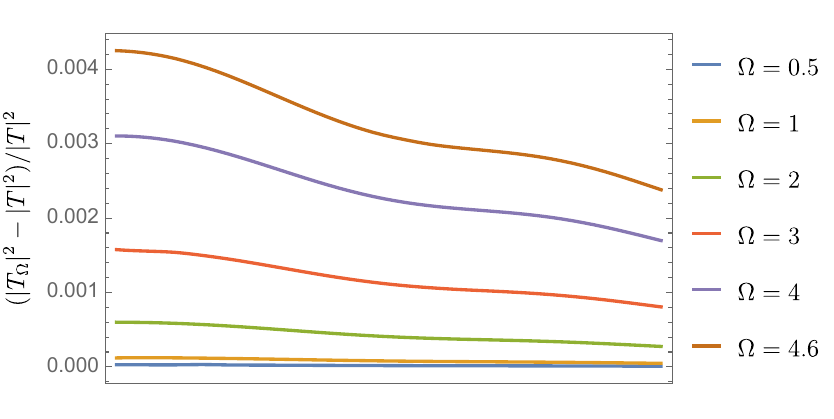}
    
    \vspace{0.45cm}
    
    \includegraphics[width=0.92\linewidth]{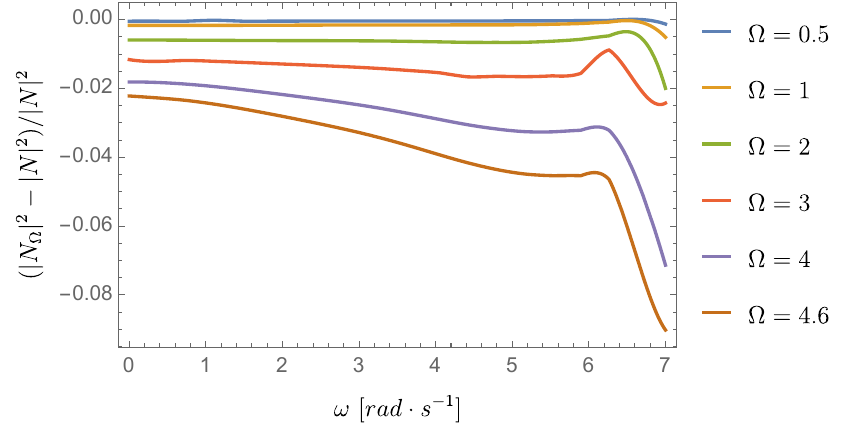}
    \caption{Percentage difference in scattering coefficients relevant to the case with no vorticity. The fluxes in the upper and lower layers have each been held constant, \emph{i.e.} $q_+=0.08\ \mathrm{m}^2/\mathrm{s}$ and $q_-=0.07\ \mathrm{m}^2/\mathrm{s}$ respectively, while the vorticity is varied from curve to curve ($\Omega$ is written in units of s$^{-1}$). We take $\Omega\le4.6$ s$^{-1}$ because for larger values the flow speed on the bottom of the channel becomes negative. We see that the maximum difference occurs for the $\left|R\right|^{2}$ coefficient, and is around 11\%. The background used for these results is shown the one shown in Fig.~\ref{fig:systemsc}.}
    \label{fig:difference}
\end{figure}

The WKB modes~(\ref{eq:wkb}) become exact in the limit of a slowly-varying background.  In this limit, it can be said that no scattering takes place, in the sense that a single incident mode evolves adiabatically into a single outgoing mode, whose wave vector is connected continuously to that of the incident mode.  The scattering amplitudes are trivial, in the sense that we have pure transmission: $\left|T\right|^{2} = 1$ for each incident mode, with all mode-mixing coefficients being zero.

Scattering is induced by the non-adiabaticity of the change of the background.  This is encoded in the fact that the WKB modes are valid wherever the background is slowly-varying, and in particular they are exact in the asymptotic regions where the background does not vary at all, and where the WKB modes just describe single plane waves.  The only region in which the WKB modes do not solve the wave equations is in the near-horizon region where the background is varying significantly.
To describe scattering, then, we can try to build an exact solution of the wave equation out of WKB modes, by introducing position-dependent amplitudes that vary only in the near-horizon region and become constant asymptotically.
(This method was also used in the appendix of~\cite{Euv__2020} to calculate scattering coefficients in the case without vorticity.)

Let us label the WKB solutions by their conserved frequency $\omega$ and a discrete label $j \in \left\{ 1,2,3 \right\}$ to distinguish between the three different solutions of the dispersion relation:
\begin{equation}
    \textbf{W}_{\omega,j}=\mathcal{N}_{\omega,j}e^{i\int^xk_{\omega,j}(x')dx'}
    \begin{pmatrix}
        \sin\theta_{\omega,j}\\
        \cos\theta_{\omega,j}
    \end{pmatrix} \,.
\end{equation}
At each position, we decompose the exact solution of the wave equations into a sum of these WKB modes:
\begin{align}
    \boldsymbol{\phi}_\omega=&\sum_j A_{\omega,j}\textbf{W}_{\omega,j} \,, \nonumber \\[0.1cm]
    \partial_x\boldsymbol{\phi}_\omega=&\sum_j ik_{\omega,j} A_{\omega,j}\textbf{W}_{\omega,j} \,.
    \label{eq:wkb_decomp}
\end{align}
Substituting Eqs.~(\ref{eq:wkb_decomp}) into the equations of motion~(\ref{eqs:mot}), we obtain a set of coupled first-order differential equations for the amplitudes $A_{\omega,j}$:
\begin{equation}
    \begin{split}
        \partial_xA_{\omega,i}=\frac{\omega^2}{2}\sum_{j\ne i}\mathcal{N}_{\omega,i}\mathcal{N}_{\omega,j}f_{ij}e^{i\int^xdx'(k_j-k_i)}A_{\omega,j}
    \end{split}
    \label{eq:aamplitudes}
\end{equation}
where
\begin{equation}
    \begin{split}
        f_{ij}=\frac{\cos\theta_{\omega,i}}{\cos\theta_{\omega,j}}&\frac{c_i}{|c_i|}\Bigg[-\frac{g}{2\Omega}\frac{\partial_x\sin^2\theta_{\omega,j}}{\sin^2\theta_{\omega,j}}+\\[0.1cm]
        &+\tan\theta_{\omega,i}\sin\theta_{\omega,j}\cos\theta_{\omega,j}\partial_x(\unotp-v_j)\Bigg].
    \end{split}
\end{equation}
It is important to note that, in the limit $\Omega \to 0$, we recover the same solutions obtained in~\cite{Euv__2020} in the absence of vorticity:
\begin{equation}
    \partial_x A_{\omega,i} = \frac{1}{2} \frac{\partial_x c}{c} e^{i \int^x dx' (k_j - k_i)} A_{\omega,j},
\end{equation}
where $i \ne j$ are the indices used to indicate the two surface waves ({\it i.e}, the co-current mode and the counter-current mode).

Equations~(\ref{eq:aamplitudes}) can be solved numerically once appropriate boundary conditions are imposed.  The counter-current modes, when traced backwards in time, are found to emanate from the horizon, and we must therefore set their amplitude to zero in the vicinity of the horizon.  The other condition to be met is that, in the upstream asymptotic region, only one incident wave (either the co-current or the vorticity wave) is present, which means that the other has zero amplitude there.  In a final step, we normalize the amplitude of the single ingoing wave to 1; then, because the amplitudes multiply the WKB modes (which are asymptotically normalized), the asymptotic values of the amplitudes are exactly the corresponding scattering amplitudes.

While only the scattering coefficients relating an incident co-current mode to outgoing co- and counter-current modes -- more precisely, the $T$, $R$ and $N$ coefficients -- are directly comparable to their counterparts in the case without vorticity, it is instructive to do so in order to gain an appreciation for how significantly the presence of vorticity affects the scattering process.
In Fig.~\ref{fig:difference}, we show the percentage difference in these particular scattering coefficients for the cases with and without vorticity. For the purpose of this comparison, the total flux is fixed in each layer: with vorticity, it has constant values in both the upper and lower layers, and the sum of these is equal to the flux used in the case without vorticity. 
These show that the scattering amplitudes are relatively robust to the introduction of the vorticity layer; the most affected is the $R$ coefficient, with $\left|R\right|^{2}$ undergoing a shift of $\sim 10 \%$ for the largest value of $\Omega$ considered.

\end{document}